\documentclass{article}
\usepackage{amssymb}
\usepackage{amsfonts}
\usepackage{amsmath}
\usepackage{graphicx}
\usepackage{mathrsfs}

\begin{document}
\noindent Published in \textit{Oxford Handbook in Philosophy of Physics,} R.
Batterman (ed.), Oxford University Press, 2013.\bigskip
\begin{center}
\bigskip

{\LARGE\textbf{INDISTINGUISHABILITY} }
\bigskip

{\Large SIMON SAUNDERS}
\end{center}

\bigskip

\noindent By the end of the 19th century the concept of particle
indistinguishability had entered physics in two apparently quite independent
ways: in statistical mechanics, where, according to Gibbs, it was needed in
order to define an extensive entropy function; and in the theory of
black-body radiation, where, according to Planck, it was needed to
interpolate between the high frequency (Wien law) limit of thermal radiative
equilibrium, and the low frequency (Rayleigh-Jeans) limit. The latter, of
course, also required the quantization of energy, and the introduction of
Planck's constant:\ the birth of quantum mechanics.

It was not only quantum mechanics. Planck's work, and later that of Einstein
and Debye, foreshadowed the first quantum field theory as written down by
Dirac in 1927. Indistinguishability is essential to the interpretation of
quantum fields in terms of particles (Fock space representations), and
thereby to the entire framework of high-energy particle physics as a theory
of local interacting fields.

In this essay, however, we confine ourselves to particle
indistinguishability in low energy theories, in quantum and classical
statistical mechanics describing ordinary matter. We are also interested in
indistinguishability as a \textit{symmetry}, to be treated in a uniform way
with other symmetries of physical theories, especially with space-time
symmetries. That adds to the need to study permutation symmetry in classical
theory -- and returns us to Gibbs and the derivation of the entropy function.

The concept of particle indistinguishability thus construed faces some
obvious challenges. It remains controversial, now for more than a century,
whether classical particles \textit{can} be treated as indistinguishable; or
if they can, whether the puzzles raised by Gibbs are thereby solved or
alleviated; and if so, how the differences between quantum and classical
statistics are to be explained. The bulk of this essay is on these
questions. In part they are philosophical. As Quine remarked:

\begin{quotation}
\noindent Those results [in quantum statistics] seem to show that there is
no difference even in principle between saying of two elementary particles
of a given kind that they are in the respective places $a$ and $b$ and that
they are oppositely placed, in $b$ and $a$. It would seem then not merely
that elementary particles are unlike bodies; it would seem that there are no
such denizens of space-time at all, and that we should speak of places $a$
and $b$ merely as being in certain states, indeed the same state, rather
than as being occupied by two things. (Quine 1990, 35).
\end{quotation}

\noindent He was speaking of indistinguishable particles in quantum
mechanics, but if particles in classical theory are treated the same way,
the same questions arise. \smallskip

\noindent This essay is organized in three sections. The first is on the
Gibbs paradox and is largely expository. The second is on particle
indistinguishability, and the explanation of quantum statistics granted that
classical particles just like quantum particles can be treated as
permutable. The third is on the more philosophical questions raised by
sections 1 and 2, and on the question posed by Quine. There is a special
difficulty in matters of ontology in quantum mechanics, if only because of
the measurement problem.\footnote{%
See Wallace (2013), Bacciagaluppi (2013).} I shall, so far as is
possible, be neutral on this this. My conclusions apply to most realist
solutions of the measurement problem, and even some non-realist ones.
\bigskip 
\bigskip

\begin{center}
{\Large\textbf{1. THE GIBBS PARADOX}}
\end{center}
\bigskip

\noindent {\large\textbf{1.1 Indistinguishability and the quantum }}\bigskip 

\noindent Quantum theory began with a puzzle over the statistical
equilibrium of radiation with matter. Specifically, Planck was led to a
certain combinatorial problem: for each frequency $\upsilon _{s}$, what is
the number of ways of distributing an integral number $N_{s}$ of `energy
elements' over a system of $C_{s}$ states (or `resonators')?

\begin{quotation}
\noindent The distribution of energy over each type of resonator must now be
considered, first, the distribution of the energy $E_{s}$ over the $C_{s}$
resonators with frequency $\upsilon _{s}$. If $E_{s}$ \ is regarded as
infinitely divisible, an infinite number of different distributions is
possible. We, however, consider - and this is the essential point - $E_{s}$
to be composed of a determinate number of equal finite parts and employ in
their determination the natural constant $h$= 6.55$\times 10^{-27}$ erg sec.
This constant, multiplied by the frequency, $\upsilon _{s}$, of the
resonator yields the energy element \ $\Delta \epsilon _{s}$ in ergs, and
dividing $E_{s}$ by $h\upsilon _{s}$, we obtain the number $N_{s}$, of
energy elements to be distributed over the $C_{s}$ resonators. (Planck 1900, 239).\footnote{%
I have used a different notation from Planck's for consistency with the
notation in the sequel.}
\end{quotation}

\noindent Thus was made what is quite possibly the most successful single
conjecture in the entire history of physics:\ the existence of Planck's
constant $h$, postulated in 1900 in the role of energy quantization.

The number of distributions $Z_{s}$, or \textit{microstates} as we shall
call them, as a function of frequency, was sought by Planck in an effort to
apply Boltzmann's statistical method to calculate the energy-density $%
\overline{E_{s}}$ of radiative equilibrium as a function of temperature $T$
and of $Z_{s}$. To obtain agreement with experiment he found
\begin{equation}
Z_{s}=\frac{(N_{s}+C_{s}-1)!}{N_{s}!(C_{s}-1)!}.
\end{equation}%
The expression has a ready interpretation: it is \textit{the number of ways
of distributing }$N_{s}$\textit{\ indistinguishable elements over }$C_{s}$%
\textit{\ distinguishable cells} -- of noting only \textit{how many}
elements are in which cell, not \textit{which} element is in which cell.%
\footnote{%
A\ microstate as just defined can be specified as a string of $N_{s}$
symbols `$p$' and $C_{s}-1$ symbols `$|$' (thus, for e.g. $N_{s}=3$, $%
C_{s}=4 $, the string $p||pp|$ corresponds to one particle in the first
cell, none in the second, two in the third, and none in the fourth). The
number of distinct strings is $(N_{s}+C_{s}-1)!$ divided by $%
(C_{s}-1)!N_{s}! $, because permutations of the symbol `$|$' among
themselves or the symbol `$p$' among themselves give the same string. (This
derivation of (1)\ was given by Ehrenfest in 1912.)} Equivalently, the
microstates are \textit{distributions invariant under permutations}. When
this condition is met, we call the elements \textit{permutable}.\footnote{%
I take `indistinguishable' and `permutable' to mean the same. But others
take `indistinguishable' to have a broader meaning, so I will give up that
word and use `permutable' instead.} Following standard physics terminology,
they are \textit{identical} if these elements, independent of their
microstates, have exactly the same properties (like charge, mass, and spin).

Planck's `energy elements' at a given frequency were certainly identical;
but whether or not it followed that they\textit{\ should} be considered
permutable was hotly disputed. Once interpreted as particles (`light
quanta'), as Einstein proposed, there was a natural alternative: why not
count microstates as distinct if they differ in which particle is located in
which cell, as had Boltzmann in the case of material particles? On that
count the number of distinct microstates should be:
\begin{equation}
Z_{s}=C_{s}^{N_{s}}.
\end{equation}%
Considered in probabilistic terms, again as Einstein proposed, if each of
the $N_{s}$ elements is assigned one of the $C_{s}$ cells at random,
independent of each other, the number of such assignments will be given by
(2), each of them equiprobable.

But whilst (2) gave the correct behaviour for $\overline{E_{s}}$ in the
high-frequency limit (Wien's law), it departed sharply from the Planck
distribution at low frequencies. Eq.(1) was empirically correct, not (2).
The implication was that if light was made of particles labelled by
frequency, they were particles that could not be considered as independent
of each other at low frequencies.\footnote{%
The \textit{locus classicus} for this story is Jammer [1966], but see also
Darrigol [1991].}

Eq.(1) is true of bosons; bosons are represented by totally symmetrized
states in quantum mechanics and quantum field theory; totally symmetrized
states are entangled states. There is no doubt that Einstein, and later Schr%
\"{o}dinger, were puzzled by the lack of independence of light-quanta at low
frequencies. They were also puzzled by quantum non-locality and
entanglement. It is tempting to view all these puzzles as related.\footnote{%
Or as at bottome the same, as argued most prominently by Howard [1990].}
Others concluded that light could not after all be made of particles, or
that it is made up of both particles and waves, or it is made up of a
special category of entities that are not really objects at all.\footnote{%
As suggested by Quine. See French and Kraus [2006] for a comprehensive
survey of debates of this kind.} We shall come back to these questions
separately.

For Planck's own views on the matter, they were perhaps closest to Gibbs'.%
\footnote{%
See Planck [1912], [1921] and, for commentary, Rosenfeld [1959].} Gibbs had
arrived at the concept of particle indistinguishability quite independent of
quantum theory. To understand this development, however, considerably more
stage-setting is needed, in both classical statistical mechanics and
thermodynamics, the business of sections 1.2-1.4. (Those familiar with the
Gibbs paradox may skip directly to 1.5.)
\bigskip
\bigskip

\noindent{\large\textbf{1.2 The Gibbs paradox in thermodynamics}}\bigskip \nolinebreak 

\noindent Consider the entropy of a volume $V$ of gas composed of $N_{A}$
molecules of kind $A$ and $N_{B}$ molecules of kind $B$\footnote{%
This section largely follows van Kampen [1984].}. It differs from the
entropy of a gas at the same temperature and pressure when $A$ and $B$ are
identical. The difference is:
\begin{equation}
-kN_{A}\log N_{A}-kN_{B}\log N_{B}+k(N_{A}+N_{B})\log (N_{A}+N_{B})
\end{equation}%
where $k$ is Boltzmann's constant, $k=1.38$ $\times 10^{-16\text{ }}$erg $%
K^{-1}$. The expression (3) is unchanged no matter how similar $A$ and $B$
are, even when in practise the two gases cannot be distinguished; but it
must vanish when $A$ and $B$ are the same. This is the Gibbs paradox in
thermodynamics.

It is not clear that the puzzle as stated is really paradoxical, but it
certainly bears on the notion of identity -- and on whether identity admits
of degrees. Thus Denbigh and Redhead argue:

\begin{quotation}
\noindent The entropy of mixing has the same value...however alike are the
two substances, but suddenly collapses to zero when they are the same. It is
the absence of any `warning' of the impending catastrophe, as the substances
are made more and more similar, which is the truly paradoxical feature
(Denbigh and Redhead [1989, 284].)
\end{quotation}

\noindent The difficulty is more severe for those who see thermodynamics as
founded on operational concepts. Identity, as distinct from similarity under
all practical measurements, seems to outstrip any possible experimental
determination.

To see how experiment does bear on the matter, recall that the classical
thermodynamic entropy is an \textit{extensive} function of the mass (or
particle number) and volume. That is to say, for real numbers $\lambda $,
the thermodynamic entropy $S$ as a function of $N$ and $V$ scales linearly:
\begin{equation*}
S(\lambda N,\lambda V,T)=\lambda S(N,V,T),\text{ }\lambda \in \mathbb{R}.%
\text{ }
\end{equation*}%
By contrast the pressure and temperature are \textit{intensive} variables
that do not scale with mass and volume. The thermodynamic entropy function
for an ideal gas is:
\begin{equation}
S(P,T,N)=\frac{5}{2}Nk\log T-Nk\log P+cN
\end{equation}%
where $c$ is an arbitrary constant. It is extensive by inspection.

The extensivity of the entropy allows one to define the analogue of a
density -- entropy per unit mass or unit volume \ -- important to
non-equilibrium thermodynamics, but the concept clearly has its limits:\ for
example, it is hardly expected to apply to gravitating systems, and more
generally ignores surface effects and other sources of inhomogeneity.
 It is
to be sharply distinguished from \textit{additivity} of the entropy, needed
to define a total entropy for a collection of equilibrium systems each
separately described -- typically, as (at least initially) physically
isolated systems. The assumption of additivity is that a total entropy can
be defined as their sum:
\begin{equation*}
S_{A+B}=S_{A}(N_{A},V_{A},T_{A})+S_{B}(N_{B},V_{B},T_{B}).\text{ }
\end{equation*}%
It is doubtful that any \textit{general} statement of the second law would
be possible without additivity. Thus, collect together a dozen equilibrium
systems, some samples of gas, others homogeneous fluids or material bodies,
initially isolated, and determine the entropy of each as a function of its
temperature, volume, and mass. Energetically isolate them from external
influences, but allow them to interact with each other in any way you like
(mechanical, thermal, chemical, nuclear), so long as the result is a new
collection of equilibrium systems. Then the second law can be expressed as:
the sum of the entropies of the latter systems is equal to or greater than
the sum of the entropies of the former systems.\footnote{%
See Lieb and Yngvason [1999] for a statement of the second law at  this  level of generality.}

Now for the connection with the Gibbs paradox. The thermodynamic entropy
difference between states $1$ and $2$ is defined as the integral, over any 
\textit{reversible} process\footnote{%
Meaning a process which at any point in its progress can be reversed, to as
good an approximation as is required. Necessary conditions are that
temperature gradiants are small and effects due to friction and turbulence
are small (but it is doubtful these are sufficient).} that links the two
states, of $dQ/T$, that is as the quantity:
\begin{equation*}
\Delta S=S_{2}-S_{1}=\int_{1}^{2}\frac{dQ}{T}
\end{equation*}%
where $dQ$ is the heat transfer. If the insertion or removal of a partition
between $A$ and $B$ is to count as a reversible process, then from
additivity and given that negligible work is done on the partition it
follows there will be no change in entropy, so no entropy of mixing. This
implies the entropy must be extensive. Conversely extensivity, under the
same presupposition, and again given additivity, implies there is no entropy
of mixing.

Whether or not the removal of a partition between $A$ and $B$ \textit{should}
count as a reversible process is another matter: surely not if means are
available to tell the two gases apart. Thus if a membrane is opaque to $A$,
transparent to $B$, under compression work $P_{A}dV$ must be done against
the partial pressure $P_{A}$ in voiding one part of the cylinder of gas $A$
(and similarly for $B$), where:
\begin{equation*}
P_{A}=\frac{N_{A}}{N_{A}+N_{B}}P,\text{ }P_{B}=\frac{N_{B}}{N_{A}+N_{B}}P.
\end{equation*}%
The work $dW$ required to separate the two gases isothermally at temperature 
$T$ is related to the entropy change and the heat transfer by: 
\begin{equation*}
dQ=dW+TdS.
\end{equation*}%
Using the equation of state for the ideal gas to determine $dW=PdV$
\begin{equation*}
PV=kNT
\end{equation*}%
where $N=N_{A}+N_{B}$, the result is the entropy of mixing, Eq.(3). However,
there can be no such semi-permeable membrane when the two gases are
identical,\footnote{%
At least in the absence of Maxwell demons: see section 3.1.} 

Would it matter to the latter conclusion if the differences between the two
gases were sufficiently small (were ignored or remained undiscovered)?
But as van Kampen argues, it is hard to see how the chemist will be led into
any \textit{practical }error in ignoring an entropy of mixing, if he cannot
take mechanical advantage of it. Most thermodynamic substances, in practise,
are composites of two or more substances (typically, different isotopes),
but such mixtures are usually treated as homogeneous. In thermodynamics, as
a science based on operational concepts, the meaning of the entropy function
does not extend beyond the competencies of the experimenter:
\begin{quotation}
\noindent Thus, whether such a process is reversible or not depends on how
discriminating the observer is. The expression for the entropy depends on
whether or not he is able and willing to distinguish between the molecules $%
A $ and $B.$ This is a paradox only for those who attach more physical
reality to the entropy than is implied by its definition. (Van Kampen [1984, 307].)
\end{quotation}

\noindent A similar resolution of the Gibbs paradox was given by Jaynes
[1992]. It appears, on this reading, that the entropy is not a real physical
property of a thermodynamic system, independent of our knowledge of it.
According to Van Kampen, it is attributed to a system on the basis of a
system of \textit{conventions} -- on whether the removal of a partition is
to be counted as a reversible process, and on whether or not the entropy
function for the two samples of gas is counted as extensive. That explains
why the entropy of mixing is an all-or-nothing affair.
\bigskip
\bigskip
\newpage
\noindent {\large\textbf{1.3 The Gibbs paradox in statistical mechanics}}

\bigskip

\noindent Thermodynamics is the one fundamental theory of physics that might
lay claim to being based on operational concepts and definitions. The
situation is different in statistical mechanics, where the concept of
entropy is not limited to equilibrium states, nor bound to the concept of
reversibility.

There is an immediate difficulty, however; for the classical derivations of
the entropy in statistical mechanics yield a function that is \textit{not}
extensive, even as an idealization. That is, classically, there is \textit{%
always} an entropy of mixing, even for samples of the same gas. If the
original Gibbs paradox was that there was no entropy of mixing in the limit
of identity, the new paradox is that there is.\footnote{%
I owe this turn of phrase to Jos Uffink.}

To see the nature of the problem, it will suffice to consider the ideal gas,
using the Boltzmann definition of entropy, so-called\footnote{%
Boltzmann defined the entropy in several different ways; see Bach [1990].}.
The state of a system of $N$ particles is represented by a set of $N$ points
in the $6-$dimensional 1-particle phase space (or $\mu -$\textit{space}), or
equivalently, by a single point in the total $6N-$dimensional phase space $%
\Gamma ^{N}$. A\textit{\ fine-graining} of $\Gamma ^{N}$ is a division of
this space into cells of equal volume $\tau ^{N}$ (corresponding to a
division of $\mu -$space into cells of volume $\tau $, where $\tau $ has
dimensions of $[$momentum$]^{3}[$length$]^{3}$)$.$ A \textit{coarse-graining}
is a division of $\Gamma ^{N}$ into regions with a given range of energy.
For weakly interacting particles these regions can be parametrized by the
one-particle energies $\epsilon _{s}$, with $N_{s}$ the number of particles
with energy in the range $[\epsilon _{s},\epsilon _{s}+\Delta \epsilon _{s}]$%
, and the coarse-graining extended to $\mu -$space as well. These numbers
must satisfy:
\begin{equation}
\sum_sN_{s}=N;\text{ }\sum_sN_{s}\epsilon _{s}=E
\end{equation}%
where $E$ is the total energy. Thus, for any fine-grained description
(microstate) of the gas, which specifies how, for each $s$, $N_{s}$
particles are distributed over the fine-graining, there is a definite
coarse-grained description (macrostate) which only specifies the number in
each energy range. Each macrostate corresponds to a definite volume of phase
space.

We can now define the Boltzmann entropy of a gas of $N$ particles in a given
microstate: \textit{it is proportional to the logarithm of the volume, in }$%
\Gamma ^{N}$\textit{, of the corresponding macrostate}$.$ In this the choice
of $\tau $ only effects an additive constant, irrelevant to entropy
differences.

This entropy is computed as follows. For each $s$, let there be $C_{s}$
cells in $\mu -$space of volume $\tau $ bounded by the energies $\epsilon
_{s},\epsilon _{s}+\Delta \epsilon ,\ $containing $N_{s}$ particles.
Counting microstates as distinct if they differ in which particles are in
which cells, we use (2)\ for the number of microstates, each with the same
phase space volume $\tau ^{N_{s}}$, yielding the volume:
\begin{equation}
Z_{s}\tau ^{N_{s}}=C_{s}^{N_{s}}\tau ^{N_{s}}.
\end{equation}
\noindent The product of these quantities (over $s$) is the $N-$particle
phase-space volume of the macrostate $N_{1},N_{2},...,N_{s},..$ for just 
\textit{one} way of partitioning the $N$ particles among the various $1-$%
particle energies. There are
\begin{equation}
\frac{N!}{N_{1}!..N_{s}!....}
\end{equation}%
partitionings in all. The total phase space volume $W^{B}$ (`$B$' is for
Boltzmann) of the macrostate $N_{1},N_{2},...,N_{s},..$ is the product of
terms (6) (over $s$) and (7):%
\begin{equation}
W^{B}=\frac{N!}{N_{1}!..N_{s}!....}\prod_{s}C_{s}^{N_{s}}\tau
^{N_{s}}
\end{equation}%
and the entropy is:
\begin{equation*}
S^{B}=k\log W^{B}=k\log \left[ \frac{N!}{N_{1}!..N_{s}!....}%
\prod_{s}C_{s}^{N_{s}}\tau ^{N_{s}}\right] .
\end{equation*}%
From the Stirling approximation for $x$ large, $\log x!\approx x\log x-x$:
\begin{equation}
S^{B}\approx kN\log N+k\sum_{s}N_{s}\log \frac{C_{s}}{N_{s}}+kN\log \tau .
\end{equation}%
By inspection, this entropy function is not extensive. When the spatial
volume and particle number are doubled, the second and third expressions on
the RHS scale properly, but not the first. This picks up a term $kN\log 2$,
corresponding to the $2^{N}$ choices as to which of the two sub-volumes
contains which particle.

One way to obtain an extensive entropy function is to simply subtract the
term $kN\log N$. In the Stirling approximation (up to a constant scaling
with $N$ and $V$) that is equivalent to dividing the volume (8)\ by $N!.$
But with what justification? If, after all, permutations of particles did
not yield distinct fine-grained distributions, the factor (7) would not be
divided by $N!$; it would be set equal to unity. \noindent Call this \textit{%
the }$N!$\textit{\ problem}. This is itself sometimes called the Gibbs
paradox, but is clearly only a fragment of it. It is the main topic of
sections 1.5 and 2.1.\bigskip \bigskip

\noindent{\large\textbf{1.4 The equilibrium entropy}}\bigskip

\noindent Although not needed in the sequel, for completeness we obtain the
equilibrium entropy, thus making the connection with observable quantities.%
\footnote{%
For a text-book derivation using our notation, see e.g. Hercus [1950].}

A system is in equilibrium when the entropy of its coarse-grained
distribution is a maximum; that is, when the entropy is stationary under
variation of the numbers $N_{s}\rightarrow N_{s}+\delta N_{s}$, consistent
with (5), i.e. from (9):
\begin{equation}
0=\delta \overline{S}^{B}=\sum_{s}[\delta N_{s}\log C_{s}-\delta N_{s}\log
N_{s}-\delta N_{s}]
\end{equation}%
where
\begin{equation}
\sum_s\delta N_{s}=0;\text{ }\sum_s\delta N_{s}\epsilon
_{s}=0.
\end{equation}%
If the variations $\delta N_{s}$ were entirely independent, each term in the
summand (10) would have to vanish. Instead introduce Lagrange multipliers $a$%
, $\beta $ for the respective constraint equations (11). Conclude for each $%
s $: 
\begin{equation*}
\log C_{s}-\log N_{s}-\alpha -\beta \epsilon _{s}=0.
\end{equation*}%
Rearranging:%
\begin{equation}
N_{s}=C_{s}(e^{-\alpha -\beta \epsilon _{s}}).
\end{equation}%
Substituting in (9) and using (5) gives the equilibrium entropy $\overline{S}%
^{B}$:%
\begin{eqnarray}
\overline{S}^{B} &=&kN\log N+k\sum_{s}N_{s}(\alpha +\beta \epsilon
_{s})+kN\log \tau  \notag \\
&=&kN\log N+kN\alpha +k\beta E+kN\log \tau .
\end{eqnarray}%
The values of $\alpha $ and $\beta $ are fixed by (5) and (12). Replacing
the schematic label $s$ by coordinates on phase space for a monatomic gas $%
\overrightarrow{x},\overrightarrow{p}$, with $\epsilon _{s}$ the kinetic
energy $\frac{1}{2m}\overrightarrow{p}^{2}$, the sum over $N_{s}$ in the
first equation of (5) becomes:%
\begin{equation*}
e^{-\alpha }\int_{V}\int e^{-\frac{\beta }{2m}\overrightarrow{p}%
^{2}}d^{3}xd^{3}p=N.
\end{equation*}%
The spatial integral gives the volume $V$; the momentum integral gives $%
(2\pi m/\beta )^{3/2}$, so
\begin{equation*}
e^{-\alpha }=\frac{N}{V}(2\pi m/\beta )^{-3/2}.
\end{equation*}%
From the analogous normalization condition on the total energy (the second
constraint (5)), substituting (12) and given that for an ideal monatomic gas 
$E=\frac{3}{2}NkT$, deduce that $\beta =\frac{1}{kT}$. Substituting in (13),
the equilibrium entropy is:%
\begin{equation*}
\overline{S}^{B}(N,V,T)=Nk\log V+\frac{3}{2}kN\log 2\pi mkT+\frac{3}{2}%
Nk+Nk\log \tau .
\end{equation*}%
It is clearly not extensive. Compare Eq.(4), which using the equation of
state for the ideal gas takes the form (the Sackur-Tetrode equation):%
\begin{equation*}
S(N,V,T)=Nk\log \frac{V}{N}+\frac{3}{2}kN\log 2\pi mkT+cN
\end{equation*}%
where $c$ is an arbitrary constant. They differ by the term $Nk\log
N$, as already noted.\bigskip\bigskip

\noindent{\large\textbf{1.5 The $N!$ puzzle}}\bigskip 

\noindent The $N!$ puzzle is this: what justifies the subtraction of the
term $Nk\log N$ from the entropy? Or equivalently, what justifies the
division of the phase space volume Eq.(8) by $N!$?\ In fact it has a fairly
obvious answer (see section 2.1): classical particles, if identical, should
be treated as permutable, just like identical quantum particles. But this
suggestion has rarely been taken seriously.

Much more widely favoured is the view that quantum theory is needed.
Classical statistical mechanics is not after all a correct theory; quantum
statistical mechanics (Eq.(1)), in the dilute limit $C_{s}\gg N_{s},$ gives:
\begin{equation*}
Z_{s}=\frac{(N_{s}+C_{s}-1)!}{N_{s}!(C_{s}-1)!}\underset{}{\approx }\frac{%
C_{s}^{N}}{N_{s}!}
\end{equation*}%
yielding the required correction to (6)\ (setting (7) to unity). Call this
the \textit{orthodox} solution to the $N!$ puzzle.

This reasoning, so far as it goes, is perfectly sound, but it does not go
very far.\ It says nothing about \textit{why} particles in quantum theory
but not classical theory are permutable. If rationale is offered, it is that
classical particles are localized in space and hence are distinguishable (we
shall consider this in more detail in the next section); and along with
that, that the quantum state for identical particles is unchanged.\footnote{%
Statements like this can be found in almost any textbook on statistical
mechanics.} \ But how the two are connected is rarely explained.

Erwin Schr\"{o}dinger, in his book \textit{Statistical Thermodynamics}, did
give an analysis:

\begin{quotation}
\noindent It was a famous paradox pointed out for the first time by W.
Gibbs, that the same increase of entropy must not be taken into account,
when the two molecules are of the same gas, although (according to naive
gas-theoretical views) diffusion takes place then too, but unnoticeably to
us, because all the particles are alike. The modern view [of quantum
mechanics] solves this paradox by declaring that in the second case there is
no real diffusion, because exchange between like particles is not a real
event - if it were, we should have to take account of it statistically. It
has always been believed that Gibbs's paradox embodied profound thought.
That it was intimately linked up with something so important and entirely
new [as quantum mechanics] could hardly be foreseen. ((Schr\"{o}dinger [1946, 61].)
\end{quotation}

\noindent Evidently, by `exchange between like particles' Schr\"{o}dinger
meant the sort of thing that happens when gases of classical molecules
diffuse -- the trajectories of individual molecules are twisted around one
another -- in contrast to the behaviour of quantum particles, which do not
have trajectories, and so do not diffuse in this way. But as for \textit{why}
the exchange of quantum particles `is not a real event' (whereas it is
classically) is lost in the even more obscure question of what quantum
particles really are. Schr\"{o}dinger elsewhere said something more, He
wrote of indistinguishable particles as `losing their identity', as
`non-individuals', in the way of units of money in the bank (they are
`fungible'). That fitted with Planck's original idea of indistinguishable
quanta as elements of energy, rather than material things -- so, again,
quite unlike classical particles.

On this point there seems to have been wide agreement. Schr\"{o}dinger's
claims about the Gibbs paradox came under plenty of criticism, for example,
by Otto Stern, but Stern remarked at the end:

\begin{quotation}
\noindent In conclusion, it should be emphasized that in the foregoing
remarks classical statistics is considered in principle as a part of
classical mechanics which deals with individuals (Boltzmann). The conception
of atoms as particles losing their identity cannot be introduced into the
classical theory without contradiction. (Stern [1949, 534].)
\end{quotation}

\noindent This comment or similar can be found scattered throughout the
literature on the foundations of quantum statistics. \bigskip

\noindent There is a second solution to the $N!$ puzzle that goes in the
diametrically-opposite direction: it appeals only to classical theory,
precisely \textit{assuming} particle distinguishability. Call this the
`classical' solution to the puzzle.

Its origins lie in a treatment by Ehrenfest and Trkal [1920] of the
equilibrium conditions for molecules subject to disassociation into a total
of $N^{\ast }$ atoms. This number is conserved, but the number of molecules $%
N_{A}$, $N_{B},...$ formed of these atoms, of various types $A$, $B$,... may
vary. The dependence of the entropy function on $N^{\ast }$ is not needed
since this number never changes: it is the dependence on $N_{A}$ , $N_{B},...
$ that is relevant to the extensivity of the entropy for molecules of type $A
$, $B,...,$ which can be measured. By similar considerations as in section
1.3, the number of ways the $N^{\ast }$ atoms can be partitioned among $N_{A}
$ molecules of type $A$, $N_{B}$ molecules of type $B$, ... is the factor $%
N^{\ast }!/N_{A}!N_{B}!...$. This multiplies the product of all the phase
space volumes for each type of molecule, delivering the required division by 
$N_{A}!$ for molecules of type $A$, by $N_{B}!$ for molecules of type $B$,
and so on (with the dependence on $N^{\ast }$ absorbed into an overall
constant).

A similar argument was given by van Kampen [1984], but using Gibbs' methods.
The canonical ensemble for a gas of $N^{\ast }$ particles has the
probability distribution:%
\begin{equation*}
W(N^{\ast },q,p)=f(N^{\ast })e^{-\beta H(q,p)}.
\end{equation*}%
Here $(q,p$) are coordinates on the $6N^{\ast }$ dimensional phase space for
the $N^{\ast }$ particles, which we suppose are confined to a volume $%
V^{\ast }$, $H$ is the Hamiltonian, and $f$ is a normalization constant. Let
us determine the probability of finding $N$ particles with total energy $E$
in the sub-volume $V$ (so $N^{\prime }=N^{\ast }-N$ are in volume $V^{\prime
}=V^{\ast }-V$). If the interaction energy between particles in $V^{\prime }$
and $V$ is small, the Hamiltonian $H_{N^{\ast }}$ of the total system can be
approximately written as the sum $H_{N}+H_{N^{\prime }}$ of the Hamiltonians
for the two subsystems. The probability density $W(N,q,p)$ for $N$ particles
as a function of $\langle N,q,p\rangle =\langle \overrightarrow{q}_{1}%
\overrightarrow{,p}_{1};\overrightarrow{q}_{2}\overrightarrow{,p}_{2};....;%
\overrightarrow{q}_{N},\overrightarrow{p}_{N}\rangle $ where $%
\overrightarrow{q}_{i}\subset V$ \ is then the marginal on integrating out
the remaining $N^{\prime }$ particles in $V^{\prime }$, multiplied by the
number of ways of selecting $N$ particles from $N^{\ast }$ particles. The
latter is given by the binomial function:%
\begin{equation*}
\binom{N^{\ast }}{N}=\frac{N^{\ast }!}{(N^{\ast }-N)!N!}.
\end{equation*}%
The result is:
\begin{equation*}
W(N,q,p)=f(N^{\ast })\binom{N^{\ast }}{N}e^{-\beta
H_{N}(N,q,p)}\int_{V^{\prime }}e^{-\beta H_{N^{\prime }}(N^{\prime
},q^{\prime },p^{\prime })}dq^{\prime }dp^{\prime }.
\end{equation*}%
In the limit $N^{\ast }\gg N,$the binomial is to a good approximation: 
\begin{equation*}
\frac{N^{\ast }!}{(N^{\ast }-N)!N!}\approx \frac{N^{\ast N}}{N!}.
\end{equation*}%
The volume integral yields $V^{\prime N^{\ast }-N}$. For non-interacting
particles, for constant density $\rho $ $=V^{\prime }/N^{\prime }$ in the
large volume limit $V^{\prime }\gg V$ we obtain:
\begin{equation*}
W(N,q,p)\approx f(N^{\ast },V^{\ast })\frac{z^{N}}{N!}e^{-\beta H_{N}(q,p)}
\end{equation*}%
where $z$ is a function of $\rho $ and $\beta $. It has the required
division by $N!.$

Evidently this solution to the $N!$ puzzle is the same as in Ehrenfest and
Trkal's derivation: extensivity of the entropy can only be obtained for an
open system, that is, for a proper subsystem of a closed system, never for a
closed one -- and it follows precisely \textit{because} the particles are
non-permutable. The tables are thus neatly turned.\footnote{%
For another variant of the Ehrenfest-Trkall approach, see Swendsen [2002,
2006, Nagle [2004].)}.

Which of the two, the orthodox or the classical, is the `correct' solution
to the $N!$ puzzle? It is tempting to say that \textit{both} are correct,
but as answers to different questions:\ the orthodox solution is about the
thermodynamics of real gases, governed by quantum mechanics, and the
classical solution is about the consistency of a hypothetical classical
system of thermodynamics that in reality does not exist. But on either line
of reasoning, identical quantum particles are treated as radically unlike
identical classical particles (only the former are permutable).\footnote{%
Note added Sep 2016. An exception is Shigeji Fujita, who in a much-neglected article (neglected by me) argued that indistinguishability is inherited from quantum statistical mechanics in the classical limit and therefore that classical particles are permutable just as are quantum particles. See Fujita [1990].}
This fits
with the standard account of the departures of quantum from classical
statistics: they are explained by permutability. But it is a false dichotomy. 
\bigskip 
\bigskip 
\
\begin{center}
{\Large \textbf{2. INDISTINGUISHABILITY AS A UNIFORM SYMMETRY }}
\end{center}
\bigskip
\noindent {\large\textbf{2.1 Gibbs' solution}}
\bigskip 

\noindent There is another answer as to which of the two solutions to the $%
N! $ puzzle is correct: \textit{neither}. The $N!$ puzzle arises in both
classical and quantum theories and is solved in exactly the same way: by
passing to the quotient space (of phase space and Hilbert space
respectively). This is not to deny that atoms really are quantum mechanical,
or that measurements of the dependence of the entropy on particle number are
made in the way that Ehrenfest et al envisaged; it is to deny that the
combinatorics factors thus introduced are, except in special cases, either
justified or needed.

Gibbs, in his \textit{Elementary Principles in Statistical Mechanics}, put
the matter as follows:

\begin{quotation}
\noindent If two phases differ only in that certain entirely similar
particles have changed places with one another, are they to be regarded as
identical or different phases? If the particles are regarded as
indistinguishable, it seems in accordance with the spirit of the statistical
method to regard the phases as identical. (Gibbs [1902,187].)
\end{quotation}

\noindent He proposed that the phase of an $N-$particle system be unaltered
`by the exchange of places between similar particles'. Phases (points in
phase space) like this he called `generic' (and those that are altered,
`specific'). The state space of generic phases is the \textit{reduced} 
\textit{phase space} $\Gamma ^{N}/\Pi _{N}$, the quotient space under the
permutation group $\Pi _{N}$ of $N$ elements. In this space points of $%
\Gamma ^{N}$ related by permutations are identified.

The suggestion is that even classically, the expressions (6) and (7) are
wrong. (7) is replaced by unity (as already noted): there is just one way of
partitioning $N$ permutable particles among the various states so as to give 
$N_{s}$ particles to each state. But (6) is wrong too:\ it should be
replaced by the volume of \textit{reduced} phase space corresponding to the
macrostate (for $s$), the volume
\begin{equation*}
\frac{(C_{s}\tau )^{N_{s}}}{N_{s}!}.
\end{equation*}%
For the macrostate $N_{1},N_{2},...,N_{s},..$ the total reduced volume,
denote $W^{red}$ is: 
\begin{equation}
W^{red}=\prod_{s}\frac{C_{s}^{N_{s}}\tau ^{N_{s}}}{N_{s}!}=\frac{%
W^{B}}{N!}.
\end{equation}%
The derivation does not depend on the limiting behaviour of Eq.(1), or on
the assumption of equiprobability or equality of volume of each fine-grained
distribution (and is in fact in contradiction with that assumption, as we
shall see).

Given (14), there is no entropy of mixing. Consider a system of particles
all with the same energy $\epsilon _{s}$. The total entropy before mixing
is, from additivity: 
\begin{equation}
S_{A}+S_{B}=k\log \left( \frac{C_{A}^{N_{A}}\tau ^{N_{A}}}{N_{A}!}\frac{%
C_{B}^{N_{B}}\tau ^{N_{A}}}{N_{B}!}\right) .
\end{equation}%
After mixing, if $A$ and $B$ are identical:%
\begin{equation}
S_{A+B}=k\log \frac{(C_{A}+C_{B})^{N_{A}+N_{B}}\tau ^{N_{A}+N_{B}}}{%
(N_{A}+N_{B})!}.
\end{equation}%
If the pressure of the two samples is initially the same (so $%
C_{A}/N_{A}=C_{B}/N_{B})$, the quantities (15), (16) should be approximately
equal\footnote{%
Should they be exactly equal? No, because it is an \textit{additional}
constraint to insist, given that $N_{A}+N_{B}$ particles are in volume $V$ $%
_{A}+V_{B}$, that exactly $N_{A}$ are in $V_{A}$ and $N_{B}$ in $V_{B}$.} --
as can easily be verified in the Stirling approximation. But if $A$ and $B$
are not identical, and permutations of $A$ particles with $B$ particles
isn't a symmetry, we pass to the quotient spaces under $\Pi _{N_{A}}$ and $%
\Pi _{N_{B}}$ separately and take their product, and the denominator in (16)
should be $N_{A}!N_{B}!$. With that $S_{A}$ $+S_{B}$ and $S_{A+B}$ are no
longer even approximately the same.

Gibbs concluded his discussion of whether to use generic or specific phases
with the words, "The question is one to be decided in accordance with the
requirements of practical convenience in the discussion of the problems with
which we are engaged" (Gibbs [1902, 188].) \noindent Practically speaking,
if we are interested in defining an extensive classical entropy function
(even for closed systems), use of the generic phase (permutability) is
clearly desirable. On the other hand, integral and differential calculus is
simple on manifolds homeomorphic to $\mathbb{R}^{6N}$, like $\Gamma ^{N}$;
the reduced phase space $\Gamma ^{N}/\Pi _{N}$ has by contrast a much more
complex topology (a point made by Gibbs). If the needed correction, division
by $N!$, can be simply made at the end of a calculation, the second
consideration will surely trump the first.\bigskip \bigskip 

\noindent{\large\textbf{2.2. Arguments against classical indistinguishability}}%
\textit{\bigskip }

\noindent Are there principled arguments against permutability thus treated
uniformly, the same in the classical as in the quantum case? The concept of
permutability can certainly be misrepresented. Thus, classically, \textit{of
course} it makes sense to move atoms about so as to interchange one with
another, for particles have definite trajectories; in that sense an
`exchange of places' must make for a real physical difference, and in that
sense `indistinguishability' cannot apply to classical particles.

But that is not what is meant by `interchange' -- Schr\"{o}dinger was just
misleading on this point. It is interchange of points in phase space whose
significance is denied, not in configuration space over time. Points in
phase-space are in $1:1$ correspondence with the dynamically allowed
trajectories. A\ system of $N$ particles whose trajectories in $\mu -$space
swirl about one another, leading to an exchange of two or more of them in
their places in space at two different times, is described by\textit{\ each}
of $N!$ points in the $6N-$dimensional phase space $\Gamma ^{N}$, each
faithfully representing the same swirl of trajectories in $\mu -$space (but
assigning different labels to each trajectory). In passing to points of the
quotient space $\Gamma ^{N}/\Pi _{N}$ there is therefore no risk of
descriptive inadequacy in representing particle interchange in Schr\"{o}%
dinger's sense.

Another and more obscure muddle is to suppose that points of phase space can
only be identified insofar as they are all traversed by one and the same
trajectory. That appears to be the principle underlying van Kampen's
argument:

\begin{quotation}
\noindent One could add, as an aside, that the energy surface can be
partitioned in $N!$ equivalent parts, which differ from one another only by
a permutation of the molecules. The trajectory, however, does not recognize
this equivalence because it cannot jump from one point to an equivalent one.
There can be no good reason for identifying the $Z-$star [the region of
phase space picked out by given macroscopic conditions] with only one of
these equivalent parts. (Van Kampen 1984, 307).
\end{quotation}

\noindent But if the whole reason to consider the phase-space volumes of
macrostates in deriving thermodynamic behaviour is because (say by
ergodicity) they are proportional to the amount of time the system spends in
the associated macrostates, then,\textit{\ just because} the trajectory
cannot jump from one point to an equivalent one, it should enough to
consider only one of the equivalent parts of the $Z-$star. We should draw
precisely the opposite conclusion to van Kampen.

However van Kampen put the matter somewhat differently -- in terms, only, of
probability:

\begin{quotation}
\noindent Gibbs argued that, since the observer cannot distinguish between
different molecules, "it seems in accordance with the spirit of the
statistical method" to count all microscopic states that differ only by a
permutation as a single one. Actually it is exactly opposite to the basic
idea of statistical mechanics, namely that the probability of a macrostate
is given by the measure of the Z-star, i.e.\textit{\ the number of
corresponding, macroscopically indistinguishable microstates}. As
mentioned...it is impossible to justify the N! as long as one restricts
oneself to a single closed system. (van Kampen 1984, 309, emphasis added).
\end{quotation}

\noindent Moreover, he speaks of probabilities of macroscopically
indistinguishable microstates, whereas the contentious question concerns 
\textit{microscopically} indistinguishable microstates. The contentious
question is whether microstates that differ only by particle permutations,
with all physical properties unchanged -- which are in this sense
indistinguishable -- should be identified.

Alexander Bach in his book \textit{Classical Particle Indistinguishability}
defended the concept of permutability of states in classical statistical
mechanics, understood as the requirement that probability distributions over
microstates be invariant under permutations. But what he meant by this is
the invariance of functions on $\Gamma ^{N}$. As such, as probability
measures, they could never provide \textit{complete} descriptions of the
particles (unless all their coordinates coincide) -- they could not be
concentrated on individual trajectories. He called this the `deterministic
setting'. In his own words:

\begin{quotation}
\noindent \textbf{Indistinguishable Classical Particles Have No Trajectories}%
. The unconventional role of indistinguishable classical particles is best
expressed by the fact that in a deterministic setting no indistinguishable
particles exist, or - equivalently - that indistinguishable classical
particles have no trajectories. Before I give a formal proof I argue as
follows. Suppose they have trajectories, then the particles can be
identified by them and are, therefore, not indistinguishable. (Bach 1997, 7).
\end{quotation}

\noindent His formal argument was as follows. Consider the coordinates of
two particles at a given time. in 1-dimension, as an extremal of the set of
probability measures $M_{+}^{1}(\mathbb{R}^{2})$ on $\mathbb{R}^{2}$ (a $2-$%
dimensional configuration space), from which, assuming the two particles are
impenetrable, the diagonal $D=\{<x,x>\in \mathbb{R}^{2},x\in \mathbb{R}\}$
has been removed. Since indistinguishable, the state of the two particles
must be unchanged under permutations (permutability), so it must be in $%
M_{+,sym}^{1}(\mathbb{R}^{2})$, the space of symmetrized measures. It
consists of sums of delta functions of the form:
\begin{equation*}
\mu _{x,y}=\frac{1}{2}\left( \delta _{<x,y>}+\delta _{<y,x>}\right) ,\text{ }%
<x,y>\in \mathbb{R}^{2}\backslash D
\end{equation*}%
But no such state is an extremal of $M_{+}^{1}(\mathbb{R}^{2})$.

As already remarked, the argument presupposes that the coordinates of the
two particles defines a point in $M_{+}^{1}(\mathbb{R}^{2})$, the unreduced
space, rather than in $M_{+}^{1}(\mathbb{R}^{2}/\Pi _{2})$, the space of
probability measures over the reduced space $\mathbb{R}^{2}/\Pi _{2}$. In
the latter case, since $M_{+}^{1}(\mathbb{R}^{2}/\Pi _{2})$ is isomorphic to 
$M_{+,sym}^{1}(\mathbb{R}^{2})),$ there is no difficulty.\footnote{%
Bach's proof, if sound, would imply that corpuscles in the de Broglie-Bohm
pilot-wave theory are distinguishable (for discussion of particle indistinguishability in pilot-wave theory, see e.g. Brown et al (1999)).}

Bach's informal argument above is more instructive. Why not use the
trajectory of a particle to identify it, by the way it twists and turns in
space? Why not indeed:\ it that is all there is to being a particle, you
have already passed to a trajectory in the quotient space $\Gamma ^{N}/\Pi
_{N}$, for those related by permutations twist and turn in exactly the same
way. The concept of particle distinguishability is not about the trajectory
or the one-particle state: it is about the\textit{\ label} of the trajectory
or the one-particle state, or equivalently, the question of \textit{which}
particle has that trajectory, that state.
\bigskip
\bigskip

\noindent{\large\textbf{2.3\ Haecceitism}}\bigskip

\noindent Gibbs' suggestion was called `fundamentally idealistic' by
Rosenfeld, `mystical' by van Kampen, `inconsistent' by Bach; they were none
of them prepared to see in indistinguishability the rejection of what is on
first sight a purely metaphysical doctrine -- that after every describable
characteristic of a thing has been accounted for, there still remains the
question of \textit{which} thing has those characteristics.

The key word is `every'; describe a thing only partly, and the question of
which it is of several \textit{more} precisely described things is obviously
physically meaningful. But microstates, we take it, are maximal, complete
descriptions. If there is a more complete level of description it is the
microstate as given by another theory, or at a deeper level of description
by the same theory, and to the latter our considerations apply.

The doctrine, now that we have understood it correctly, has a suitably
technical name in philosophy.\ It is called \textit{haecceitism}. Its
origins are medieval if not ancient, and it was in play, one way or another,
in a connected line of argument from Newton and Clarke to Leibniz and Kant.
That centered on the need, given symmetries, including permutations, not
just for symmetry-breaking in the choice of initial conditions,\footnote{%
As in e.g. a cigar-shaped mass distribution, rather than a sphere. Of
course, this is not really a breaking of rotational symmetry, in that each
is described by relative angles and distances between masses, invariant
under rotations.} but for a choice among haecceistic differences -- in the
case of continuous symmetries, among values of absolute positions, absolute
directions, and absolute velocities. All parties to this debate agreed on
haecceitism. These choices were acts of God, with their consequences visible
only to God (Newton, Clarke); or they were humanly visible too, but in ways
that couldn't be put into words -- that could only be grasped by `intuition'
(Kant); or they involved choices \textit{not} even available to God, who can
only choose on the basis of reason; so there could be no created things such as
indistinguishable atoms or points of a featureless space (Leibniz).\footnote{%
For more on Leibniz see Saunders [2003]. For a compilation of original
sources and commentary, see Huggett [1999b].}

So much philosophical baggage raises a worry in its own right. If it is the
truth or falsity of haecceitism that is at issue, it seems unlikely that it
can be settled by any empirical finding. If that is what the extensivity of
the entropy is about, perhaps extensivity has no real physical meaning after
all. It is, perhaps, itself metaphysical -- or conventional. This was the
view advocated by Nick Huggett when he first drew the comparison between
Boltzmann's combinatorics and haecceitism.\footnote{%
Huggett [1999a], also endorsed in Albert [2000, 47-8].).}

But this point of view is only remotely tenable if haecceitism is similarly
irrelevant to empirical questions in quantum statistics. And on the face of
it that \textit{cannot} be correct. Planck was, after all, led by \textit{%
experiment} to Eq.(1). Use of the unreduced state space in quantum mechanics
rather than the reduced (symmetrized) space surely has direct empirical
consequences.

Against this two objections can be made. The first, following Reichenbach,
is that the important difference between quantum and classical systems is
the absence in quantum theory of a criterion for the re-identification of
identical particles over time. They are, for this reason, `non-individuals'
(this links with Schr\"{o}dinger's writings\footnote{%
See Schr\"{o}dinger [1984, 207-210]. The word `individual' has also
been used to mean an object answering to a unique description at a single
time (as `absolutely discernible' in the terminology of Saunders [2003],
[2006b]). Note added Sep 2016: and as re-identifiable over time (see my [2016] for further discussion}). This, rather than any failure of haecceitism, is what is
responsible for the departures from classical statistics.\footnote{%
As recently endorsed by Pooley [2006 section 8].} The second, following Post
[1970] and French and Redhead [1989] is that haecceitism must be consistent
with quantum statistics (including Planck's formula) because particles, even
given the symmetrization of the state, may nevertheless possess
`transcendental' individuality, and symmetrization of the state can itself
be understood as a dynamical constraint on the state, rather than in terms
of permutability.

Of these the second need not detain us. Perhaps metaphysical claims can be
isolated from any possible impact on physics, but better, surely, is to link
them with physics where such links are possible. Or perhaps we were just
wrong to think that haecceitism is a metaphysical doctrine: it just means
non-permutability, it is the rejection of a symmetry.

As for the first, it is simply not true that indistinguishable quantum
particles can never be re-identified over time. Such identifications are
only exact in the kinematic limit, to be sure, and even then only for a
certain class of states; but the ideal gas is commonly treated in just such
a kinematic limit, and the restriction in states applies just as much to the
reidentification of identical quantum particles that are\textit{\ not}
indistinguishable -- that are \textit{not} permutable -- but which are
otherwise entangled.

This point needs some defence. Consider first the case of non-permutable
identical particles. The $N$ particle state space is then $\mathcal{H}^{N}=%
\mathcal{H}\otimes \mathcal{H\otimes }..\mathcal{\otimes H}$, the $N-$fold
tensor product of the $1-$particle state space $\mathcal{H}$. Consider
states of the form:
\begin{equation}
|\Phi \rangle =\underset{\text{N-factors}}{\underbrace{\overset{\text{%
k-factors}}{\overbrace{|\phi \rangle _{a}\otimes |\phi _{b}\rangle \otimes
...\otimes |\phi _{c}\rangle }\otimes ...\otimes |\phi _{d}\rangle }}}
\end{equation}%
where the one-particle states are members of some orthonormal basis (we
allow for repetitions). The $k^{th}$- particle is then in the one-particle
state $|\phi _{c}\rangle $. If the particles are only weakly interacting,
and the state remains a product state, the $k^{th}-$ particle can also be
assigned a one-particle state at later times, namely the unitary evolute of $%
|\phi _{c}\rangle $. Even if more than one particle has the initial state $%
|\phi _{c}\rangle $, still it will be the case that each particle in that
state has a definite orbit under the unitary evolution. It is true that in
those circumstances it would seem impossible to to tell the two orbits
apart, but the same will be true of two classical particles with exactly the
same representative points in $\mu -$space.\footnote{%
One might in classical mechanics add the condition that the particles are
impenetrable; but one can also, in quantum mechanics, require that no two
particles occupy the same one-particle state (the Pauli exclusion
principle). See sections 2.5, 3.3.}

Now notice the limitation of this way of speaking of particles as
one-particles states that are (at least conceptually) identifiable over
time: it does not in general apply to superpositions of states of the form
(17) -- as will naturally arise if the particles are interacting, even
starting from (17). In general, given superpositions of product states,
there is no single collection of $N$ one-particle states, or orbits of
one-particle states, sufficient for the description of the $N$ particles
over time. In these circumstances no definite histories, no orbits of
one-particle states, can be attributed to identical but distinguishable
particles either.

Now consider identical permutable quantum particles (indistinguishable
quantum particles). The state must now be invariant under permutations, so
(for vector states):
\begin{equation}
U_{\pi }|\Phi \rangle =|\Phi \rangle
\end{equation}%
for every $\pi \in \Pi ^{N}$, where $U:\pi \rightarrow U_{\pi }$ is a
unitary representation of the permutation group $\Pi ^{N}$. Given (18), $%
|\Phi \rangle $ must be of the form:
\begin{equation}
|\Phi \rangle =c\sum_{\pi \in \Pi _{N}}|\phi _{\pi (a)}\rangle
\otimes |\phi _{\pi (b)}\rangle \otimes ...\otimes |\phi _{\pi (c)}\rangle
\otimes ...\otimes |\phi _{\pi (d)}\rangle
\end{equation}%
and superpositions thereof. Here $c$ is a normalization constant, $\pi \in
\Pi _{N}$ is a permutation of the $N$ symbols $\{a,b,...,c,..d\}$ (which
again, may have repetitions), and as before, the one-particle states are
drawn from some orthonormal set in $\mathcal{H}$. If non-interacting, and
initially in the state (19), the particle in the state $|\phi _{c}\rangle $
can \textit{still} be reidentified over time -- as the particle in the state
which is the unitary evolute of $|\phi _{c}\rangle $.\footnote{%
As we shall see, there is a complication in the case of fermions (section 3.3),
although it does not effect the point about identity over time.} That is to
say,\textit{\ for entanglements like this}, one-particle states can still be
tracked over time. It is true that we can no longer refer to the state as
that of the $k^{th}$ particle, in contrast to states of the form (17), but
that labelling -- unless shorthand for something else, say a lattice
position -- never had any physical meaning. As for more entangled states --
for superpositions of states of the form (19) -- there is of course a
difficulty; but it is the same difficulty as we encountered for identical
but distinguishable particles.

Reichenbach was therefore right to say that quantum theory poses special
problems for the reidentification of identical particles over time, and that
these problems derive from entanglement; but not from the `mild'\footnote{%
The terminology is due to Penrose [2004, 598]. See Ghirardi et al [2002], 2004] for the claim that entanglement due to (anti)-symmetrization isn't
really entanglement at all} form of entanglement required by
symmetrization itself (as involved in states of the form (19)), of the sort
that explains quantum statistics. On the other hand, this much is also true:
permutability does rule out appeal to the reduced density matrix to
distinguish each particle in time (defined, for the $k^{th}$ particle, by
taking the partial trace of the state over the Hilbert space of all the
particles save the $k^{th}$). Given (anti)symmetrization, the reduced
density matrices will all be the same. But it is hard to see how the reduced
density matrix can provide an operational as opposed to a conceptual
criterion for the reidentification of an individual system over time.

What would an operational criterion look like? here is a simple example: a
helium atom in the canister of gas by the laboratory door is thereby
distinguished from one in the high-vacuum chamber in the corner, a criterion
that is preserved over time. This means: the one-particle state localized in
the canister is distinguished from the one in the vacuum chamber.

We shall encounter this idea of reference and reidentification by location
(or more generally by properties) again, so let us give them a name:\ call
it `individuating reference', and the properties concerned `individuating
properties'. In quantum mechanics the latter can be represented in the usual
way by projection operators. Thus if $P_{can}$ is the projector onto the
region of space $\Delta _{can}$ occupied by the canister, and $P_{cham}$
onto the region $\Delta _{cham}$ occupied by the vacuum chamber, and if $%
|\chi _{1}\rangle ,|\chi _{2}\rangle $ are localized in $\Delta _{can}$ (and
similarly $|\psi _{1}\rangle ,|\psi _{2}\rangle $ in $\Delta _{cham}$ ),
then even in the superposition (where $|c_{1}|^{2}+|c_{2}|^{2}=1$)
\begin{eqnarray*}
|\Phi \rangle &=&c_{1}\frac{1}{\sqrt{2}}\left( |\chi _{1}\rangle \otimes
|\psi _{1}\rangle +|\psi _{1}\rangle \otimes |\chi _{1}\rangle \right) \\
&&+c_{2}\frac{1}{\sqrt{2}}\left( |\chi _{2}\rangle \otimes |\psi _{2}\rangle
+|\psi _{2}\rangle \otimes |\chi _{2}\rangle \right)
\end{eqnarray*}%
one can still say there is a a state in which one particle is in region $%
\Delta _{can}$ and one in $\Delta _{cham}$ (but we cannot say which); still
we have:%
\begin{equation}
(P_{can}\otimes P_{cham}+P_{cham}\mathbb{\otimes }P_{can})|\Phi \rangle
=|\Phi \rangle .
\end{equation}%
If the canister and vacuum chamber are well-sealed, this condition will be
preserved over time. Individuating properties can be defined in this way
just as well for permutable as for non-permutable identical particles.
\bigskip

\noindent It is time to take stock. We asked whether the notion of
permutability can be applied to classical statistical mechanics. We found
that it can, in a way that yields the desired properties of the statistical
mechanical entropy function, bringing it in line with the classical
thermodynamic entropy. We saw that arguments for the unintelligibility of
classical permutability in the literature are invalid or unsound, amounting,
at best, to appeal to the philosophical doctrine of haecceitism. We knew
from the beginning that state-descriptions in the quantum case should be
invariant under permutations, and that this has empirical consequences, so
on the most straight-forward reading of haecceitism the doctrine is false in
that context. Unless it is emasculated from all relevance to physics,
haecceitism cannot be true a priori. We wondered if it was required or
implied if particles are to be reidentified over time, and found the answer
was no to both, in the quantum as in the classical case. We conclude:
permutation symmetry holds of identical classical particles just as it does
of identical quantum particles, and may be treated in the same way, by
passing to the quotient space.

Yet an important lacuna remains, for among the desirable consequences of
permutation symmetry in the case of quantum particles are the departures
from classical statistics -- statistics that are unchanged in the case of
classical particles. Why is there this difference? \bigskip\ 

\bigskip

\noindent{\large\textbf{2.4 The explanation of quantum statistics}}\bigskip 

\noindent Consider again the classical reduced phase-space volume for the
macrostate $N_{1},N_{2},...,N_{s},..$, as given by Eq.(14):%
\begin{equation}
W^{red}=\prod_{s}\frac{C_{s}^{N_{s}}\tau ^{N_{s}}}{N_{s}!}.
\end{equation}%
In effect, Planck replaced the one-particle phase-space volume element $\tau 
$, hitherto arbitrary, by $h^{3}$, and changed the the factor $Z_{s}$ by
which it was multiplied to obtain:%
\begin{equation}
W^{BE}=\prod_{s}\frac{(N_{s}+C_{s}-1)!h^{3N_{s}}}{N_{s}!(C_{s}-1)!}.
\end{equation}%
Continuing from this point, using the method of sections 1.3 and 1.4 one is
led to the equilibrium entropy function and equation of state for the ideal
Bose-Einstein gas. The entire difference between this and the classical
ideal gas is that for each $s$, the integer $C_{s}^{N}$ is replaced by $%
(N_{s}+C_{s}-1)!/(C_{s}-1)!.$ What is the rational for this? It does not
come from particle indistinguishability (permutability); that has already
been taken into account in (21).

Let us focus on just one value of $s$, that is, on $N_{s}$ particles
distributed over $C_{s}$ cells, all of the same energy (and hereafter drop
the subscript $s$). At the level of the fine-grained description, in term of
how many (indistinguishable) particles are in each (distinguishable) cell, a
microstate is specified by a sequence of fine-grained occupation numbers $%
<n_{1}$, $n_{2}$, ... , $n_{C}>$, where $\sum_{j=1}^{C}n_{j}=N;$ there are
many such corresponding to the coarse-grained description ($N$, $C$) (for a
single value of $s$). Their sum is
\begin{equation}
\sum_{\substack{ \text{all sequences }<n_{1},..,n_{C}>\text{ }  \\ 
\text{s.t. }\sum_{k=1}^{C}n_{k}=N}}1\text{ }=\text{ }\frac{(N+C-1)!}{N!(C-1)!%
}
\end{equation}%
as before. But here is another mathematical identity:\footnote{%
A special case of the multinomial theorem (see e.g. Rapp [1972, 49-50]).}
\begin{equation}
\sum_{\substack{ \text{all sequences }<n_{1},..,n_{C}>  \\ \text{%
s.t. }\sum_{k=1}^{C}n_{k}=N}}\frac{1}{n_{1}!...n_{C}!}\text{ }=\text{ }\frac{%
C^{N}}{N!}.
\end{equation}%
In other words, the difference between the two expressions (21) and (22),
apart from the replacement of the unit $\tau $ by $h^{3}$, is that \textit{%
in quantum mechanics every microstate }$<n_{1}$\textit{, }$n_{2}$\textit{,
... , }$n_{C}>$\textit{\ has equal weight, whereas in classical mechanics
each is weighted by the factor} $(n_{1}!...n_{C}!)^{-1}$.

Because of this weighting, a classical\ fine-grained distribution where the $%
N$ particles are evenly distributed over the $C$ cells has a much greater
weight than one where most of the particles are concentrated in a small
handful. In contrast, in quantum mechanics, the weights are always the same.
Given that `weight', one way or another, translates into statistics,
particles weighted classically thus tend to repel, in comparison to their
quantum mechanical counterparts; or put the other way, quantum particles
tend to bunch together, in comparison to their classical counterparts.
\begin{figure}[h, !]
    \includegraphics [scale=0.4]{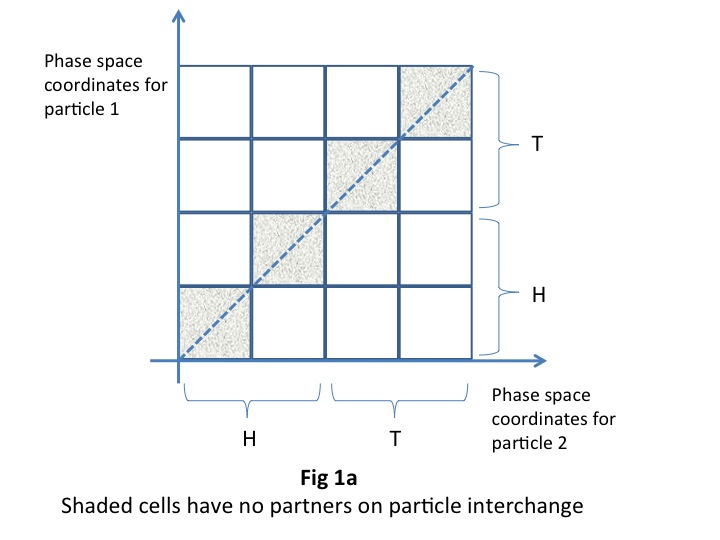}
\end{figure}
\begin{figure}[h, !]
    \includegraphics[scale=0.4]{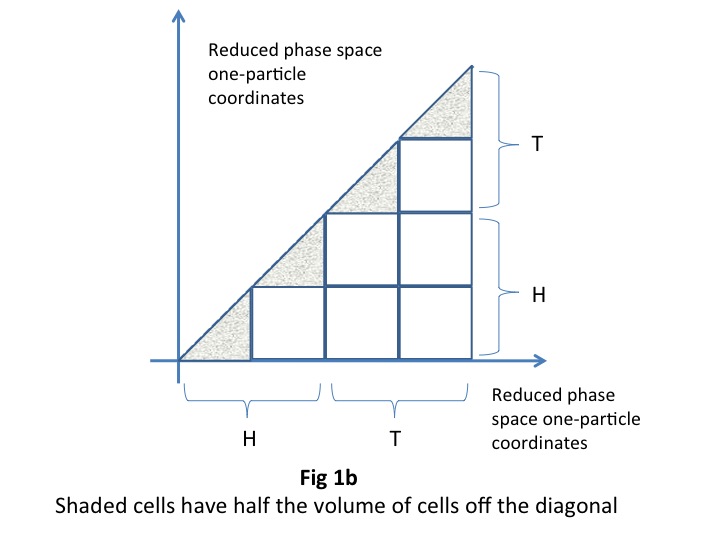}
\end{figure}

That is what the weighting does, but why is it there? Consider fig.1a, a representation of phase space
for $N=2$, $C=4$. Suppose, for concreteness, we are modelling two classical,
non-permutable identical coins, such that the first two cells correspond to
one of the coins landing heads ($H$), and the remainder to that coin landing
tails ($T$) (and similarly for the other coin).\footnote{%
Of course for macroscopic coins, the assumption of degeneracy of the energy
is wildly unrealistic, but let that pass.} The cells along the diagonal
correspond not just to both coins landing heads or both landing tails --
they are cells in which the two coins have \textit{all }their fine-grained
properties the same. For any cell away from the diagonal, there is a
corresponding cell that differs only in which coin has which fine-grained
property (its reflection in the diagonal). Their combined volume in phase
space is therefore twice that of any cell on the diagonal.

The same is true in the reduced phase space, fig.1b. For $N=3$ there are
three such diagonals; cells along these have one half the volume of the
others. And there is an additional boundary, where all three particles have
the same fine-grained properties, each with one sixth their volume. The
weights in Eq.(24) follow from the structure of reduced phase space, as
faithfully preserving ratios of volumes of microstates in the unreduced
space. As explained by Huggett [1999], two classical identical coins, if
permutable, still yield a weight for $\{H,T\}$ twice that of the weight for $%
\{H,H\}$ or $\{T,T\}$, just as for non-permutable coins, that is with
probabilities one-half, one-quarter, and one-quarter respectively.

Contrast quantum mechanics, where subspaces of Hilbert space replace regions
of phase space, and subspace dimension replaces volume measure. Phase space
structure, insofar as it can be defined in quantum theory, is derivative and
emergent. Since the only measure available is subspace dimension, each of a
set of orthogonal directions in each subspace is weighted precisely the same
-- yielding, for the symmetrized Hilbert space, Eq.(23) instead.\footnote{%
One way of putting this is that in the quantum case, the measure on phase
space must be discrete, concentrated on points representing each unit cell
of `volume' $h^{3}$. For early arguments to this effect see Planck [1912),
Poincar\'{e} [1911, 1912].}

But there are two cases when subspace dimension and volume measure are
proportional to one another -- or rather, for we take quantum theory as
fundamental, for when phase-space structure,complete with volume measure,
emerges from quantum theory\footnote{%
See Wallace (2013)}. One is in the limit $C\gg N$, when the
contribution from the states along the diagonals is negligible in comparison
to the total (fig. 2b), and the other is when the full Hilbert space for
non-permutable particles is used. That is why permutability makes a
difference to statistics in the quantum case but not the classical: for $%
N\approx C$, as in fig 2a, the dimensionality measure departs significantly
from the volume measure (in fig 2a, as five-eighths to one-half). For $N=2$, 
$C=2$ there are just three orthogonal microstates, each of equal weight.
Take two two-state quantum particles (qubits) as quantum coins, and the
probabilities $\{H,H\}$, $\{T,T\}$, $\{H,T\}$ are all one-third.
  \begin{figure}[h,!]
  \includegraphics[scale=0.4]{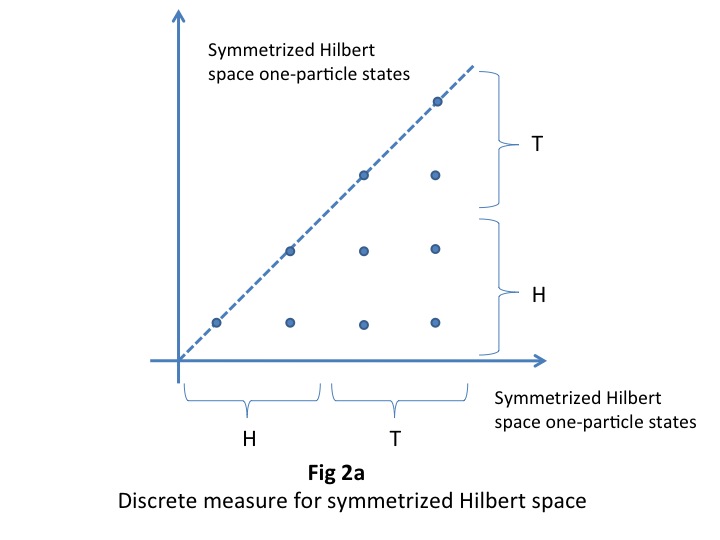}
\end{figure}
\begin{figure}[h,!]
    \includegraphics[ scale=0.4]{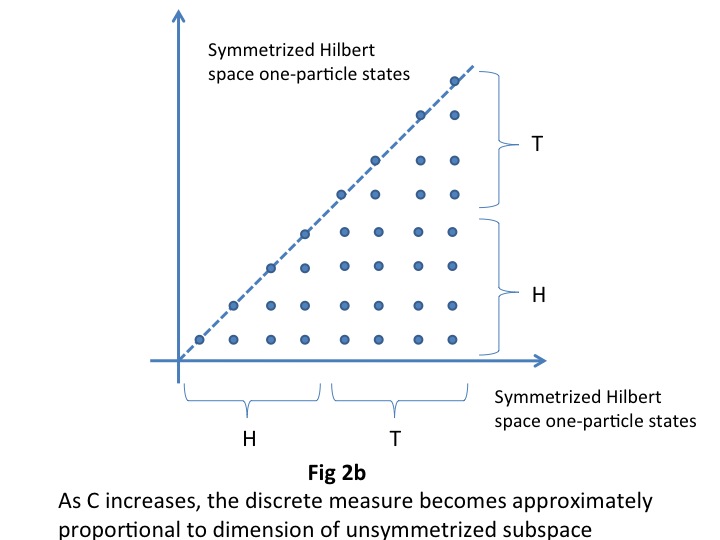}
\end{figure}

Is there a remaining puzzle about quantum statistics -- say, the
non-independence of permutable quantum particles, as noted by Einstein?
Statistical independence fails, in that the state cannot be specified for $%
N-1$ particles, independent of the state of the $N^{th}$, but that is true
of classical states on reduced phase space too (or, indeed, for
permutation-invariant states on the unreduced phase space -- see Bach
[1997]). Find a way to impose a discrete measure on a classical permutable
system, and one can hope to reproduce quantum statistics as well (Gottesmann
[2005]). Quantum holism has some role to play in the explanation of quantum
statistics, but like entanglement and identity over time, less than meets
the eye.\bigskip \bigskip

\noindent{\large\textbf{2.5 Fermions}}\bigskip

\noindent We have made almost no mention so far of \ fermions. In fact most
of our discussion applies to fermions too, but there are some differences.

Why are there fermions at all? The reason is that microstates in quantum
theory are actually rays, not vector states $|\phi _{c}\rangle $, that is,
they are $1-$dimensional subspaces of Hilbert space. As such they are
invariant under multiplication by complex numbers of unit norm. If only the
ray need be invariant under permutations, there is an alternative to
Eq.(18), namely: 
\begin{equation}
U_{\pi }|\Psi ^{FD}\rangle =e^{i\theta }|\Psi ^{FD}\rangle
\end{equation}%
where $\theta \in \lbrack 0,2\pi ]$. Since any permutation can be decomposed
as a product of permutations $\pi _{ij}$ (that interchange $i$ and $j$),
even or odd in number, and since $\pi _{ij}\pi _{ij}=\mathbb{I}$, it follows
that (18) need not be obeyed after all: there is the new possibility that $%
\theta =0$ or $\pi $ for even and odd permutations respectively. Such states
are \textit{antisymmetrized}, i.e. of the form:
\begin{equation}
|\Psi ^{FD}\rangle =\frac{1}{\sqrt{N!}}\sum_{\pi \in \Pi _{N}}\text{%
sgn}(\pi )|\phi _{\pi (a)}\rangle \otimes |\phi _{\pi (b)}\rangle \otimes
...\otimes |\phi _{\pi (c)}\rangle \otimes ...\otimes |\phi _{\pi (d)}\rangle
\end{equation}%
where sgn($\pi $)$=1$ ($-1$) for even (odd) permutations, and superpositions
thereof.

An immediate consequence is that, unlike in (19), every one-particle state
in (26) must now be orthogonal to every other: repetitions would
automatically cancel, leaving no contribution to $|\Psi ^{FD}\rangle $.
Since superpositions of states (19) with (26) satisfy neither (18) or (25),
permutable particles in quantum mechanics must be of one kind or the other.%
\footnote{%
This is to rule out parastatistics -- representations of the permutation
group which are not one-dimensional (see e.g. Greiner and M\"{u}ller
[1994]). This would be desirable (since parastatistics have not been
observed, except in 2-dimensions, where special considerations apply), but I
doubt that it has really been explained.}

The connection between phase space structure and antisymmetrization of the
state is made by the Pauli exclusion principle -- the principle that no two
fermions can share the same complete set of quantum numbers, or
equivalently, have the same one-particle state. In view of the effective
identification of elementary phase space cells of volume $h^{3}$ with rays
in Hilbert space, fermions will be constrained so that no two occupy the
same elementary volume. In other words, in terms of microstates in phase
space, the $n_{k}$'s are all zeros or ones. In place of Eq.(23), we obtain
for the number of microstates for the coarse-grained distribution $\langle
C,N\rangle $ (as before, for a single energy level $s$):
\begin{equation}
\sum_{\substack{ \text{fine grainings }n_{k}\in \lbrack 0,1] \\ 
\text{s.t. }\sum_{k=1}^{C}n_{k}=N}}1=\frac{C!}{(C-N)!N!}.
\end{equation}%
Use of (27) in place of (1) yields the entropy and equation of state for the
Fermi-Dirac ideal gas. It is, of course, extensive. A classical phase space
structure emerges from this theory in the same limit $C\gg N$ (for each $s$%
)\ as for the Bose-Einstein gas, when the classical weights for cells along
the diagonals are small in comparison to the total. Away from this limit,
whereas for bosons their weight is too small (as suppressed by the factor $%
(n_{1}!...n_{C},)^{-1}$), for fermions their weight is too large (as not
suppressed enough; they should be set equal to zero). Thus fermions tend to
repel, in comparison to non-permutable particles.\footnote{%
The situation is a little more complicated, as antisymmetry in the spin part
of the overall state forces symmetry in the spatial part - which can lead to
spatial bunching (this is the origin of the homopolar bond in quantum
chemistry).}\bigskip \bigskip 

\begin{center}
{\Large\textbf{3 ONTOLOGY}}
\end{center}
\bigskip 
\noindent The explanation of quantum statistics completes the main argument of this essay: permutation symmetry falls in place as with
any other exact symmetry in physics, and applies just as much to classical
systems of equations that display it as to quantum systems.\footnote{%
But see Belot (2013) for pitfalls in defining such symmetries.}
In both cases \textit{only quantities invariant under permutations are
physically real}. This is the sense in which `exchange between like
particles is not a real event'; it has nothing to do with the swirling of
particles around each other, it has only to do with haecceistic redundancies
in the mathematical description of such particles, swirling or otherwise.
Similar comments apply to other symmetries in physics, where instead of
haecceistic differences one usually speaks of coordinate-dependent
distinctions.

In both classical and quantum theory state-spaces can be defined in terms
only of invariant quantities. In quantum mechanics particle labels need
never be introduced at all (the so-called `occupation number formalism') --
a formulation recommended by Teller [1995]. Why introduce quantities
(particle labels) only to deprive them of physical significance? What is
their point if they are permutable? We come back to Quine's question and to
eliminativism.

There are two sides to this question. One is whether, or how, permutable
particles can be adequate as ontology (section 3.1), and link in a
reasonable way with philosophical theories of ontology (sections 3.2 and
3.3). The other question is whether some other way of talking might not be
preferable, in which permutability as a symmetry does not even arise
(section 3.4). \bigskip \bigskip\ 

\noindent{\large\textbf{3.1 The Gibbs paradox, again}}\bigskip 

\noindent A first pass at the question of whether permutable entities are
really objects is to ask how they may give rise to non-permutable objects.
That returns us to the Gibbs paradox in the sense of section 1.2: How
similar do objects have to be to count as identical?

On this problem (as opposed to the $N!$ problem) section 2 may seem a
disappointment. It focused on indistinguishability as a symmetry, but the
existence of a symmetry (or otherwise)\ seems just as much an all-or-nothing
affair as identity. But section 2 did more than that:\ it offered a
microscopic dynamical analysis of the process of mixing of two gasses.

In fact, not even the $N!$ problem is entirely solved, for we would still
like to have an extensive entropy function even where particles are
obviously non-identical, say in the statistical behaviour of large objects
(like stars), and of small but complex objects like fatty molecules in
colloids.\footnote{%
This problem afflicts the orthodox solution to the Gibbs paradox, too (and
was raised as such by Swendsen [2006]).} In these cases we can appeal to the
Ehrenfest-Trkaal-van Kampen approach, but only given that we can arrive at a
description of such objects as distinguishable: how do we do that, exactly?

The two problems are related, and an answer to both lies in the idea of
individuating properties, already introduced, and the idea of phase-space
structure as emergent, already mentioned. For if particles (or bound states
of particles) acquire some dynamically stable properties, there is no reason
that they should not play much the same role, in the definition of effective
phase-space structure, as do intrinsic ones. Thus two or more non-identical
gases may arise, even though their elementary constituents are identical and
permutable, if all the molecules of one gas have some characteristic
arrangement, different from those of the other. The two gases will be
non-identical only at an effective, emergent level of description to be
sure, and permutation symmetries will still apply at the level of the full
phase-space. The effective theory will have only approximate validity, in
regimes where those individuating properties are stable in time. Similar
comments apply to Hilbert-space structures.\footnote{%
That is, the familiar intrinsic properties of particles (like charge, spin
and mass) may be state-dependent: String theory and supersymmetric theories
provide obvious examples. See Goldstein et al [2005a,b] for the argument
that all particles may be treated as permutable, identical or otherwise.}
\begin{figure}[h]
    \includegraphics[trim=-1cm 0cm 0cm 0cm, clip=true,scale=0.4]{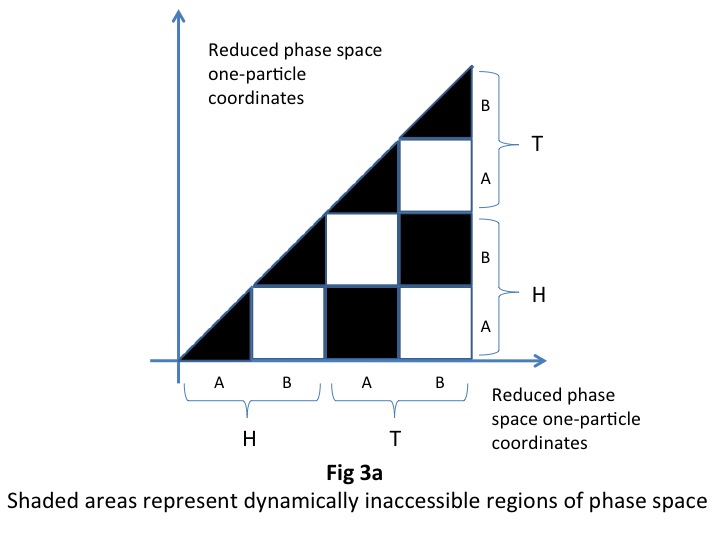}
\end{figure}
\begin{figure}[h!]
    \includegraphics[trim=0cm 4cm 0cm 3cm, clip=true, scale=0.4]{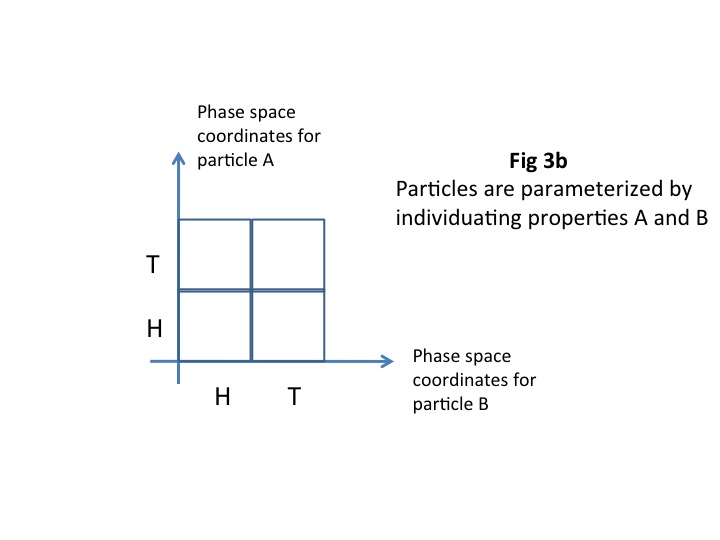}
\end{figure}
In illustration, consider again figure1b for two classical permutable
coins. Suppose that the dynamics is such that one of the coins always lands
on top of the other. Their gravitational potential energy is therefore
different.\footnote{%
That makes it harder to maintain the fiction of degeneracy of the energy,
but let that pass too.} This fact is recorded in the microstate:\ each coin
not only lands either heads ($H$) or tails ($T$), but lands either above ($A$%
)\ or below ($B$). It follows that certain regions of the reduced phase
space are no longer accessible, among them the cells on the diagonal for
which all the properties of the two coins are the same (shaded, figure 3a). By inspection, the available phase space has the effective structure
of an unreduced phase space for\textit{\ distinguishable} coins, the $A$
coin and the $B$ coin (figure 3b). It is tempting to add `even if there
is no fact of the matter as to which of the coins is the $A$ coin, and which
is the $B$ coin', but there is another way of putting it: the coin which is
the $A$ coin is the one rotating one way, the $B$ coin is the one rotating the other way.\footnote{%
For further discussion, see section 3.3. Whether the $A$ coin after one toss is
the same as the $A$ coin on another toss (and likewise the $B$ coin) will
make a difference to the effective dynamics.}

The elimination of the diagonals makes no difference to particle statistics
(since this is classical theory), but analogous reasoning applies to the
quantum case, where it does. Two quantum coins (qubits), thus dynamically
distinguished, will land one head and one tail with probability one half,
not one third.
\begin{figure}[h!]
    \includegraphics[trim=-1cm 2cm 2cm 2cm, clip=true, scale=0.4]{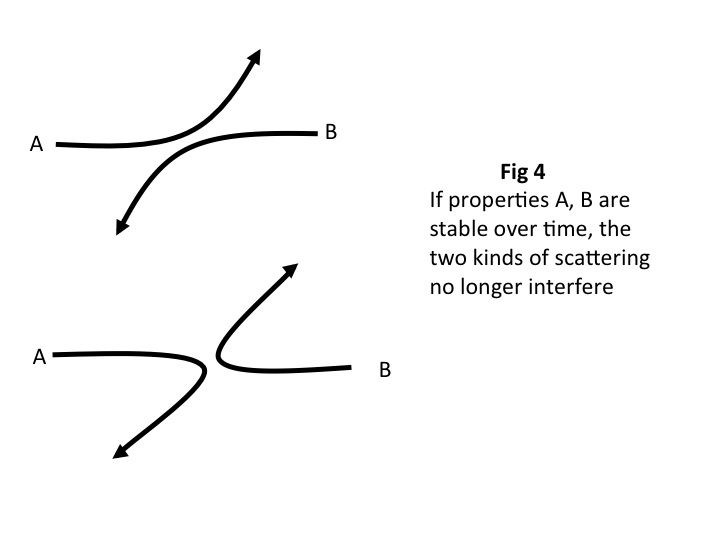}
\end{figure}

The argument carries over unchanged in the language of Feynman diagrams.
Thus the two scattering processes depicted in figure 4 cannot (normally)\ be
dynamically distinguished if the particles are permutable Correspondingly,
there is an interference effect that leads to a difference in the
probability distributions for scattering processes involving permutable
particles from those for distinguishable particles. But if dynamical
distinctions $A$ and $B$ can be made between the two particles, stable over
time (in our terms, if $A$ and $B$ are individuating properties), the
interference terms vanish, and the scattering amplitudes will be the same as
for distinguishable particles.

The same procedure can be applied to $N=N_{A}+N_{B}$ coins, $N_{A}$ of which
land above and $N_{B}$ land below. The result for large $N_{A},$ $N_{B}$ is
an effective phase space representation for two non-identical gases $A$ and $%
B$, each separately permutable, each with an extensive entropy function,
with an entropy of mixing as given by (3). And it is clear this
representation admits of degrees: it is an effective representation, more or
less accurate, more or less adequate to practical purposes.

But by these means we are a long way from arriving at an effective phase
space theory of $N$ distinguishable particles. That would require, at a
minimum, $N$ distinct individuating properties of the kind we have described
-- at which point, if used in an effective phase space representation, the
original permutation symmetry will have completely disappeared. But it is
hardly plausible (for microscopic systems), when $N$ is large, that a
representation like this can be dynamically defined. Even where there are
such individuating properties, as with stars and (perhaps) with colloids, it
is hard to see what purposes their introduction would serve -- their
dynamical definition -- unless it is to model explicitly a Maxwell demon.%
\footnote{%
The memory records of such a demon in effect provide a system of
individuating properties for the $N$ particles.} On this point we are in
agreement with van Kampen. But it must be added:\ we do better to recognize
that the use of unreduced phase space, and the structure $\mathbb{R}^{6N}$
underlying it, is in general, and at best, a mathematical simplification,
introducing distinctions in thought that are not instantiated in the
dynamics.

That seems to be exactly what Gibbs thought on the matter. He had, \ recall,
an epistemological argument for passing to reduced phase space -- that
nothing but similarity in qualities could be used to identify particles
across members of an ensemble of gasses -- but he immediately went on to say:

\begin{quotation}
\noindent And this would be true, if the ensemble of systems had a
simultaneous objective existence. \textit{But it hardly applies to the
creations of the imagination}. In the cases which we have been
considering\ldots .it is not only possible to conceive of the motion of an
ensemble of similar systems simply as possible cases of the motion of a
single system, but it is actually in large measure for the sake of
representing more clearly the possible cases of the motion of a single
system that we use the conception of an ensemble of systems. The perfect
similarity of several particles of a system will not in the least interfere
with the identification of a particular particle in one case with a
particular particle in another. (Gibbs [1902, 188], emphasis added.)
\end{quotation}

\noindent If pressed, it may be added that a mathematician can always
construct a domain of objects in set theory, or in one-one correspondence
with the real numbers, each number uniquely represented.\footnote{%
For further discussion, see Muller and Saunders [2008]. (Set-theory of
course yields rigid structures \textit{par excellence}.)} Likewise for
reference to elements of non-rigid structures, which admit non-trivial
symmetries -- for example, to a particular one of the two roots of $-1$ in
the complex number field, or to a particular orientation on $3-$dimensional
Euclidean space, the left-handed orientation rather than the right-handed
one.\footnote{%
This was also, of course, a key problem for Kant. For further discussion,
and an analysis of the status of mirror symmetry given parity violation in
weak-interaction physics, see Saunders [2007].} But it is another matter
entirely as to whether reference like this, in the absence of individuating
properties, can carry over to physical objects. The whole of this essay
can be seen as an investigation of whether it can in the case of the concept
of particle; our conclusion is negative.

The lesson may well be more general. It may be objects in mathematics are
always objects of singular thought, involving, perhaps, an irreducible
indexical element. If, as structuralists like Russell and Ramsey argued, the
most one can hope for in representation of physical objects is structural
isomorphisms with objects of direct acquaintance, these indexical elements
can be of no use to physics. It is the opposite conclusion to
Kant's.\bigskip\ \bigskip\

\noindent{\large\textbf{3.2 Philosophical logic}}\bigskip 

\noindent A second pass at our question of whether permutable entities can
be considered as objects is to ask whether they can be quantified over in
standard logical terms. Posed in this way, the question takes us to language
and objects as values of bound variables. Arguably, the notion of object has
no other home; physical theories are not directly about objects, properties,
and identity in the logical sense (namely equality).

But if are to introduce a formal language, we should be clear on its limits.
We are not trying to reproduce the mathematical workings of a physical
theory in its terms. That would hardly be an ambitious, but hardly novel
undertaking; it is the one proposed by Hilbert and Russell, that so inspired
Carnap and others in the early days of logical empiricism. Our proposal is
more modest. The suggestion is that by formalization we gain clarity on the 
\textit{ontology} of a physical theory, not rigour or clarity of deduction
-- or even of explanation. But it is ontology subject to symmetries: in our
case, permutability. We earlier saw how invariant descriptions and invariant
states (under the permutation group) suffice for statistical mechanics,
suffice even for the description of individual trajectories; we should now
see how this invariance is to be cashed out in formal, logical terms.

Permutability of objects, as a symmetry, has a simple formal expression:
predicates should be invariant (have the same truth value) under
permutations of values of variables. Call such a predicate `totally
symmetric'.

Restriction to predicates like these certainly \textit{seems} onerous. Thus
take the simple case where there are only two things, whereupon it is enough
for a predicate to be totally symmetric that it be symmetric in the usual
sense. When we say:

\begin{description}
\item[(i)] Buckbeak the hippogriff can fly higher than Pegasus the winged
horse
\end{description}

\noindent the sentence is clearly informative, at least for readers of
literature on mythical beasts; but `flies higher' is not a symmetric
predicate. How can we convey (i) without this asymmetry?

Like this: by omitting use of proper names. Let us suppose our language has
the resources to replace them with Russellean descriptions, say with
`Buckbeak-shaped' and `Pegasus-shaped' as predicates (`individuating
predicates'). We can then say in place of (i)

\begin{description}
\item[(ii)] $x$ is Buckbeak-shaped and $y$ is Pegasus-shaped and $x$ can fly
higher than $y$
\end{description}

\noindent But now (ii) gives over to the equally informative totally
symmetric predicate:

\begin{description}
\item[(iii)] $x$ is Buckbeak-shaped and $y$ is Pegasus shaped and $x$ can
fly higher than $y$, or $y$ is Buckbeak-shaped and $x$ is Pegasus-shaped and 
$y$ can fly higher than $x.$
\end{description}

\noindent The latter is invariant under permutation of $x$ and $y$.
Prefacing by existential quantifiers, it says what (i) says (modulo
uniqueness), leaving open only the question of which of the two objects is
the one that is Buckbeak-shaped, rather than Pegasus-shaped, and vice versa.
But continuing in this way -- adding further definition to the individuating
predicate -- the question that is left open is increasingly empty. If no
further specification is available, one loses nothing in referring to that
which is Buckbeak-shaped, that which is Pegasus-shaped (given that there are
just the two); or to using `Buckbeak' and `Pegasus' as mass terms, like
`butter' or `soil'. We then have from (iii):

\begin{description}
\item[(iv)] There is Buckbeak and there is Pegasus and Buckbeak can fly
higher than Pegasus, or there is Buckbeak and there is Pegasus and Buckbeak
can fly higher than Pegasus
\end{description}

\noindent With `Pegasus' and `Buckbeak' in object position, (iv) is not
permutable. We have recovered (i).

How does this work when there are several other objects? Consider the
treatment of properties as projectors in quantum mechanics. For a
one-particle projector $P$ there corresponds the $N-$fold symmetrized
projector: 
\begin{equation*}
P\otimes (I-P)\otimes ...\otimes (I-P)+(I-P)\otimes P\otimes (I-P)...\otimes
(I-P)+....+(I-P)\otimes ...\otimes (I-P)\otimes P
\end{equation*}%
where there are $N$ factors in each term of the summation, of which there
are $\binom{N}{1}=N$. For a two-particle projector of the form $P\otimes Q$,
the symmetrized operator is likewise a sum over products of projections and
their complements ($N$ factors in each), but now there will be $\binom{N}{2}%
=N(N-1)$ summands. And so on. The obvious way to mimic these constructions
in the predicate calculus, for the case of $N$ objects, is to define, for
each one-place predicate $A$, the totally symmetric $N$-ary predicate:

\begin{description}
\item[(v)] $(Ax_{1}\wedge \lnot Ax_{2}\wedge ...\wedge \lnot Ax_{N})\vee
(\lnot Ax_{1}\wedge Ax_{2}\wedge \lnot Ax_{3}\wedge ...\wedge \lnot
Ax_{N})\vee ....\vee (\lnot Ax_{1}\wedge ...\wedge \lnot Ax_{N-1}\wedge
Ax_{N}).$
\end{description}

\noindent The truth of (v) (if it is true) will not be affected by
permutations of values of the $N$ variables. It says only that exactly one
particle, or object, satisfies $A$, not which particle or object does so.
The construction starting with a two-place predicate follows similar lines;
and so on for any $n$-ary predicate for $n\leq N.$ Disjuncts of these can be
formed as well. 

Do these constructions tell us all that we need to know? Indeed they must,
given our assumption that the $N$ objects are adequately described in the
predicate calculus without use of proper names, for we have:\ 

\begin{description}
\item[Theorem 1] Let $\mathscr{L} $\ be a first-order language with
equality, without any proper names. Let $S$ be any $\mathscr{L} -$sentence
true only in models of cardinality $N$. Then there is a totally symmetric $N$%
-ary predicate $G$ $\in \mathscr{L} $ such that $\exists x_{1}...\exists
x_{N}Gx_{1}...x_{N}$ is logically equivalent to $S$.
\end{description}

\noindent (For the proof see Saunders [2006a].) Given that there is some
finite number of objects $N$, anything that can be said of them without
using proper names (with no restriction on predicates) can be said of them
using a totally symmetric $N$-ary predicate.\footnote{%
This construction was overlooked by Dieks and Lubberdink [2010] in their
criticisms of the concept of classical indistinguishable particles. They go
further, rejecting indistinguishability even in the quantum case (they
consider that particles only emerge in quantum mechanics in the limit where
Maxwell-Boltzmann statistics hold sway -- where individuating properties in
our sense can be defined).}

On the strength of this, it follows we can handle uniqueness of reference as
well, in the sense of the `that which' construction, `the unique $x$ which
is $Ax$'. In Peano's notation it is he object $\iota xAx$. Following
Russell, it is contextually defined by sentences of the form

\begin{description}
\item[(vi)] the $x$ that is an $A$ is a $B$
\end{description}

\noindent or $B(\imath x)Ax$, which is cashed out as:

\begin{description}
\item[(vii)] $\exists x(Ax\wedge \forall y(Ay\rightarrow y=x)\wedge Bx)$.
\end{description}

\noindent From Theorem 1 it follows that (vii), supplemented by information
on just how many objects there are, is logically equivalent to a sentence
that existentially quantifies over a totally symmetric predicate (like (v)).
It says that a thing which is $A$ is a $B$, that something is an $A$, and
that there are no two distinct things that are both $A$, without ever saying
which of $N$ things is the thing which is $A$.

How much of this will apply to quantum particles? All of it. Of course
definite descriptions of objects of definite number is rarely needed in talk
of atoms, and rarely available. Individuating properties at the macroscopic
level normally provide indefinite descriptions of an indeterminate number of
particles. So it was earlier; I was talking of any old helium atom in the
canister by the door, any old helium atom in the vacuum chamber, out of an
indeterminate number in each case. But sometimes numbers matter: a handful
of atoms of plutonium in the wrong part of the human body might be very bad
news indeed. Even one might be too many.

Nor need we stop with Russellean descriptions, definite or otherwise. There
are plenty of other referential devices in ordinary language that may be
significant. It is a virtue of passing from the object level, from objects
themselves (the `material mode', to use Carnap's term), to \textit{talk} of
objects (the `formal mode'), that the door is open to linguistic
investigations of quite broad scope. Still, in agreement with Carnap and
with Quine, our litmus test is compatibility with elementary logic and
quantification theory. \bigskip

\noindent To conclude: in the light of Theorem 1, and the use of
individuating properties to replace proper names, nothing is lost in passing
from non-permutable objects to permutable ones. There is no loss of
expressive content in talking of $N$ permutable things, over and above what
is lost in restricting oneself to the predicate calculus and abjuring the
use of names. That should dissipate most philosophical worries about
permutability.

There remains one possible bugbear, however, namely identity in the logical
sense (what we are calling equality). Quantum objects have long been thought
problematic on the grounds that they pose insuperable difficulties to any
reasonable account of logical equality -- for example, in terms of the
principle of identity of indiscernibles (see below). To this one can reply,\
too bad for an account of equality; the equality sign can be
taken as primitive, as is usual in formal logic.\footnote{%
See Pniower [2004] for arguments to this effect.}  (That is to say, in any
model of $\mathscr{L} $, if a language with equality, the equals sign goes
over to equality in the set-theoretic sense.) But here too one might do
better. \bigskip 
\bigskip

\noindent{\large \textbf{3.3 Identity conditions}}
\bigskip\ 

\noindent If physical theories were (among other things) directly \textit{
about} identity in the logical sense, an account of it would be available
from them. It is just because physical theories are \textit{not} like this (although
that could change) that I am suggesting the notion of object should be
formalized in linguistic terms. It is not spelt out for us directly in any
physical theory.

But by an `account of equality' I do not mean a theory of logical equality
in full generality. I mean a theory of equality only of physical objects,
and specific to a scientific language. It may better be called an account of
identity conditions, contextualized to a physical theory.

Given our linguistic methods, there is an obvious candidate: exhaustion of
predicates. That is, if $F...s..$if and only if $F...t...$, for every
predicate in $\mathscr{L} $ and for every predicate position of $F$, then $s$
and $t$ are equal. Call this $\mathscr{L} $ \textit{-equality}, denote `$%
s=_{\mathscr{L} }t$'. It is clearly a version of Leibniz's famous `principle
of identity of indiscernibles'. This is often paraphrased as the principle
that objects which share the same properties, or even the same relational
properties, are the same, but this parsing is unsatisfactory in an important
respect. It suggests that conditions of the form 
\begin{equation}
\forall y(Fsy)\leftrightarrow \forall y(Fty)\wedge \forall
y(Fys)\leftrightarrow \forall y(Fyt)
\end{equation}%
are sufficient to imply that $s$ and $t$ are equal, but more than this is
required for exhaustion of predicates. The latter also requires the truth of
sentences of the form:%
\begin{equation}
\forall y(Fsy\leftrightarrow Fty)\wedge \forall y(Fys\leftrightarrow Fyt)).
\end{equation}

\noindent These are the key to demonstrating the non-identity of many
supposed counterexamples to Leibniz's principle (of distinct objects that appear qualitatively the same; see Saunders [2003]).

$\mathscr{L} $ \textit{-}equality is the only defined notion of equality (in
first-order languages) that has been taken seriously by logicians.\footnote{%
It was first proposed by Hilbert and Bernays [1932]; it was subsequently
championed by Quine [1960], [1970].} It satisfies G\"{o}del's axioms for the
sign `$=$', used in his celebrated completeness proof for the predicate
calculus with equality, namely the axiom scheme:

\begin{description}
\item[Leibniz's law] $s=t$ $\rightarrow \bigwedge_{F\in \mathscr{L}
}\left( F..s..\leftrightarrow F..t...\right) $
\end{description}

\noindent together with the scheme $s=s$. Since one has completeness,
anything true in $\mathscr{L} $ equipped with the sign `$=$' remains true in 
$\mathscr{L} $ equipped with the sign `$s=_{\mathscr{L} }t$'. The difference
between $\mathscr{L} $ \textit{-}equality and primitive equality cannot be
stated in $\mathscr{L} $.\footnote{%
For further discussion, see Quine [1970, 61-64], and, for criticism,
Wiggins [2004, 184-88].}

But the notion that we are interested in is not $\mathscr{L} $ \textit{-}%
equality, sameness with respect to every predicate in $\mathscr{L} $ \textit{%
, }but \textit{sameness with respect to invariant predicates constructible in%
} $\mathscr{L} $, denote $\mathscr{L} ^{\ast }$. Call equality defined in
this way `physical equality', denote `$=_{\mathscr{L} ^{\ast }}$'$.$ With
that completeness is no longer guaranteed, but our concern is with ontology,
not with deduction.

In summary, we have:$\mathscr{L}$

\begin{description}
\item[physical equality] $s=_{\mathscr{L} ^{\ast }}t\underset{\text{def}}{=}%
\bigwedge_{F\in \mathscr{L} ^{\ast }}\left( Fs\leftrightarrow
Ft\right) $
\end{description}

\noindent and, as a necessary condition for physical objects (`the identity
of physical indiscernibles'):\footnote{%
For further discussion of this form of the principle of identity of
indiscernibles, see Muller and Saunders [2008, 522-23].}

\begin{description}
\item[IPI] $s=_{\mathscr{L} ^{\ast }}t\rightarrow s=t.$
\end{description}

\noindent If $s\neq _{\mathscr{L} ^{\ast }}t$, we shall say $s$ and $t$ are
`(physically) discernible'; otherwise `indiscernible'.

There are certain logical distinctions (first pointed out by Quine) for
equality in our defined sense that will prove useful. Call $s$ and $t$
`absolutely discernible' if for an open sentence $F$ in one free variable, $%
Fs$ and not $Ft$; call $s$ and $t$ `weakly discernible' (respectively
`relatively discernible') if for an open sentence $F$ in two free variables $%
Fst$ but not $Fss$ (respectively, but not $Fts$). Objects that are only
weakly or relatively discernible are discerned by failure of conditions of
the form (29), not (28).

Of these, as already mentioned, weak discernibility is of greater interest
from both a logical and physical point of view. Satisfaction of any
symmetric but irreflexive relation is enough for weak discernibility: $\neq $
and $\neq _{\mathscr{L} }$are prime examples. And many simple invariant
physical relations are symmetric and irreflexive: for example, having
non-zero relative distance in a Euclidean space (a relation invariant under
translations and rotations). Thus take Max Black's famous example of
identical iron spheres $s$, $t$, one mile apart, in an otherwise empty
Euclidean space. The spheres are weakly discerned by the relation $D$ of
being one mile apart , for if $Dst$ is true, it is not the case that $%
Dxs\leftrightarrow Dxt$ for any $x$, since $Dst$ but not $Dss$ (or $Dtt$),
so $s\neq _{\mathscr{L} ^{\ast }}t.$ And, fairly obviously, if $\mathscr{L}
^{\ast }$ contains only totally symmetric predicates, physical objects will
be at most weakly discernible.

Here as before `$s$' and `$t$' are terms, that is variables, functions of
variables, or proper names. What difference do the latter make? Names are
important to discernibility under $\mathscr{L} $-equality. Thus if it is
established that $s$ and $t$ are weakly $\mathscr{L} -$discernible, then, if
'$s$' or '$t$' are proper names, they are absolutely $\mathscr{L} -$%
discernible. In the example just given, if $Dst$ and `$s$' is a proper name,
then $Dsx$ is true of $t$ but not $s$. But the presence of names in $%
\mathscr{L} $ makes no difference to $\mathscr{L} ^{\ast }-$discernibility
(discernibility by totally symmetric predicates). Thus, even if $Dxy\in
\mathscr{L} ^{\ast },$ on entering a proper name in variable position one
does not obtain a one-place predicate in $\mathscr{L} ^{\ast }$. Permutable
objects are only weakly discernible, if discernible at all.

It remains to determine whether permutable particles\textit{\ are}
discernible at all. In the classical case, assuming particles are
impenetrable, they are always some non-zero distance apart, so the answer is
positive. Impenetrability also ensures that giving up permutability, and
passing to things which are particle states or trajectories, they will be at
least weakly discernible. Typically they will be strongly discernible, but
as Black's two spheres illustrate (supposing they just sit there), not
always.

It is the quantum case that presents the greater challenge;
indistinguishable quantum particles have long been thought to violate any
interesting formulation of Leibniz's principle of indiscernibles.\footnote{%
See\ French and Kraus [2006] for this history.} But in fact the same options
as in the classical case are there available. One can speak of that which
has such-and-such a state, or orbit, and pass to states and orbits of states
as things, giving up permutability. One-particle states or their orbits,
like classical trajectories, will in general be absolutely discernible, but
sometimes only weakly discernible -- or (failing impenetrability) not even
that. Or retaining permutability, one can speak of particles as being in one
or other states, and of $N$ particles as being in an $N-$particle state,
using only totally symmetric predicates. One then looks for a symmetric and
irreflexive relation that they satisfy.

On both strategies there is a real difficulty in the case of bosons, at
least for elementary bosons. On the first approach, there may be two bosonic
one-particle states, each exactly the same; on the second, there seems to be
no general symmetric and irreflexive relation that is always satisfied. But
the situation is different when it comes to fermions. On the first approach,
given only the mild entanglement required by antisymmetrization, one is
guaranteed that of the $N$ one-particle states, each is orthogonal to every
other, so objects as one-particle states are always absolutely discernible;
and on the second approach, again following from antisymmetrization, an
irreflexive symmetric relation can always be defined (whatever the degree of
entanglement). I shall consider them in turn.

The first strategy is not without its difficulties. To begin with, even
restricting to mildly-entangled states,\textit{\ which} one-particle states
are to be the objects replacing particles is ambiguous. The problem is
familiar from the case of the singlet state of spin: neglecting spatial
degrees of freedom the antisymmetrized state is
\begin{equation}
|\Psi _{0}\rangle =\frac{1}{\sqrt{2}}\left( |\psi _{+}^{z}\rangle \otimes
|\psi _{-}^{z}\rangle -|\psi _{-}^{z}\rangle \otimes |\psi _{+}^{z}\rangle
\right)
\end{equation}%
where $|\psi _{\pm }^{z}\rangle $ are eigenstates of spin in the $z$
direction. But this state can equally be expanded in terms of eigenstates of
spin in the $y$ direction, or of the $z$ direction: which pair of absolutely
discernible one-particle states are present, exactly?

The problem generalizes. Thus, for arbitrary orthogonal one-particle states $%
|\phi _{a}\rangle $, $|\phi _{b}\rangle $, and a two-fermion state of the
form:%
\begin{equation}
|\Phi \rangle =\frac{1}{\sqrt{2}}\left( |\phi _{a}\rangle \otimes |\phi
_{b}\rangle -|\phi _{b}\rangle \otimes |\phi _{a}\rangle \right) 
\end{equation}%
define the states (the first is just a change of notation): 
\begin{eqnarray}
|\phi _{+}^{1}\rangle  &=&|\phi _{a}\rangle ,\text{ }|\phi _{-}^{1}\rangle
_{-}=|\phi _{b}\rangle ) \\
|\phi _{+}^{2}\rangle  &=&\frac{1}{\sqrt{2}}(|\phi _{a}\rangle +|\phi
_{b}\rangle ),\text{ }|\phi _{-}^{2}\rangle _{-}=\frac{1}{\sqrt{2}}(|\phi
_{a}\rangle -|\phi _{b}\rangle )  \notag \\
|\phi _{+}^{3}\rangle  &=&\frac{1}{\sqrt{2}}(|\phi _{a}\rangle +i|\phi
_{b})\rangle ,\text{ }|\phi _{-}^{3}\rangle _{-}=\frac{1}{\sqrt{2}}(i|\phi
_{a}\rangle +|\phi _{b}\rangle ).  \notag
\end{eqnarray}%
They yield a representation of the rotation group. One then has, just as for
components of spin:%
\begin{eqnarray*}
|\Phi \rangle  &=&\frac{1}{\sqrt{2}}(|\phi _{+}^{1}\rangle |\otimes |\phi
_{-}^{1}\rangle -|\phi _{-}^{1}\rangle |\otimes |\phi _{+}^{1}\rangle  \\
&=&\frac{1}{\sqrt{2}}(|\phi _{+}^{2}\rangle |\otimes |\phi _{-}^{2}\rangle
-|\phi _{-}^{2}\rangle |\otimes |\phi _{+}^{2}\rangle =\frac{1}{\sqrt{2}}%
(|\phi _{+}^{3}\rangle |\otimes |\phi _{-}^{3}\rangle -|\phi _{-}^{3}\rangle
|\otimes |\phi _{+}^{3}\rangle 
\end{eqnarray*}%
and an ambiguity in attributing one-particle states to the two particles
arises with (31) as with (30). I shall come back to this in section 3.4.

This difficulty can be sidestepped at the level of permutable particles,
however. In the case of (30), we may weakly discern the particles by the
relation `opposite spin', with respect to any direction in space (Saunders
[2003], [2006b], Muller and Saunders [2008]). Thus if $\sigma ^{x}$, $\sigma
^{y}$, $\sigma ^{z}$ are the Pauli spin matrices, the self-adjoint operator 
\begin{equation}
\sigma ^{x}\otimes \sigma ^{x}=\sigma ^{y}\otimes \sigma ^{y}=\sigma
^{z}\otimes \sigma ^{z}
\end{equation}%
has eigenvalue $-1$ in the singlet state $|\Psi _{0}\rangle $ , with the
clear interpretation that the spins are anticorrelated (with respect to any
direction in space). Asserting this relation does not pick out any direction
in space, no more than saying Black's spheres are one-mile apart picks out
any position in space.

For the construction in the generalized sense (32), define projection
operators onto the states $|\phi _{\pm }^{k}\rangle $
\begin{equation*}
P_{\pm }^{k}=P_{|\phi _{\pm }^{k}\rangle },\text{ }k=1,2,3
\end{equation*}%
and define the self-adjoint operators: 
\begin{equation*}
(P_{+}^{k}-P_{-}^{k})\otimes (P_{+}^{k}-P_{-}^{k}),\text{ }k=1,2,3.
\end{equation*}%
Each has eigenvalue $-1$ for $|\Phi \rangle $, and likewise picks out no
`direction' in space (i.e the analogue of (33)\ is satisfied). Moreover, one
can define sums of such in the case of finite superpositions of states of
the form (31), by means of which fermions can be weakly discerned.

On the strength of this, one can hope to weakly discern bosons that are
composites of fermions, like helium atoms. And even in the case of
elementary bosons, self-adjoint operators representing irreflexive,
symmetric relations required of any pair of bosons have been proposed.%
\footnote{%
See Muller and Seevink [2009]. Their idea was to use certain commutator
relations that could not be satisfied were there only a single particle.}.
The difficulty of reconciling particle indistinguishability in quantum
mechanics with the IPI looks well on its way to being solved.\bigskip\bigskip

\noindent{\large\textbf{3.4 Eliminativism.}} \bigskip 

\noindent We are finally in a position to assemble the arguments for and
against eliminativism -- that is, for and against renouncing talk of
permutable objects in favour of non-permutable objects defined in terms of
individuating properties, whether points in $\mu -$space, trajectories,
one-particle states, or orbits of one-particle states. The gain, \textit{%
usually}, is absolute discernibility. On the other hand we have found that
quantification over permutable objects satisfies every conservative
guideline we have been able to extract from elementary logic (with the
possible exception of identity conditions for elementary bosons). And there
remains another conservative guideline:\ we should maintain standard
linguistic usage where possible.

That stacks the odds against eliminativism, for talk of particles, and not
just of one-particle states, is everywhere in physics. But even putting this
to one side eliminativism would seem to fare poorly, for (anti)symmetrized
states are generically entangled, whereupon no set of $N$ one-particle
states will suffice for the description of $N$ particles. And where such a
set is available, given sufficiently mild entanglements, it may be
non-unique.

Against this there are two objections. The first is that we anyway know the
particle concept is stretched to breaking point in strongly-interacting
regimes. There the best we can say is that there are quantum fields, and,
perhaps, superpositions of states of different particle number. Where the
latter can be defined, one can talk of modes of quantum fields instead. In
the free-field limit, or as defined by a second-quantization of a particle
theory,\footnote{%
For a discussion of the relation between second quantized and free-field
theories (fermionic and bosonic respectively), see Saunders [1991], [1992]}
such modes are in one-one correspondence with one-particle states (or, in
terms of Fourier expansions of the fields, in correspondence with
`generalized' momentum eigenstates). The elimination of particles in favour
of fields and modes of fields is thus independently motivated.

The second objection is that we cannot lightly accept indeterminateness in
attributing a definite set of $N$ one-particle states to an $N-$particle
system, for it applies equally to particles identified by individuating
properties. That is, not even the property of being a bound electron in a
helium atom in the canister by the corner, and being one in the vacuum
chamber by the door, hold unambiguously. The construction (32) applies just
as much to (the antisymmetric version of) (20).\footnote{%
Note added Sep 2016. This difficulty was pointed out in Ghirardi et al [2002, 84-86], but it was dismissed on the grounds that measurements involving the `wrong' choice of states `are extremely difficult to perform and of no practical interest' (p.86).}

But\textit{\ this} difficulty we recognize as a fragment of the measurement
problem. Specifically, it is the `preferred basis problem': into what states
does a macroscopic superposition collapse (if there is any collapse)? -- or,
if macroscopic superpositions exist: what singles out the basis in which
they are written? Whatever settles this question (decoherence, say) will
dictate the choice of basis used to express the state in terms of
macroscopic individuating properties.footnote{%

Whether such a choice of basis -- or such a solution to the preferred basis
problem -- can extend to a preferred basis at the microscopic level is moot.
It depends, to some extent, on the nature of the solution (decoherence only
goes down so far). Of course it is standard practise in quantum theory to
express microscopic states in terms of a basis associated with
physically-interpreted operators (typically generators of one-parameter
spacetime symmetry groups, or in terms of the dynamical quantities that are
measured). The use of quantum numbers for bound states of electrons in the
atom, for energy, orbital angular momentum, and components of angular
momentum and spin -- in conventional notation, quadruples of numbers $%
\langle n,l,m_{l},m_{s}\rangle $ -- is a case in point. When energy
degeneracies are completely removed (introducing an orientation in space)
one can assign these numbers uniquely. The Pauli exclusion principle then
dictates that every electron has a unique set of quantum numbers. Use such
quadruples as names and talk of permutable particles can be eliminated.
 
It is now clearer that the first objection adds support to the second.
Quadruples of quantum numbers provide a natural replacement for particles in
atoms; modes of quantum fields (and their excitation numbers) provide a
natural replacement for particles involved in scattering. And in
strongly-interacting regimes, even modes of quantum fields give out (or they
have only a shadow existence, as with virtual particles). All this is as it
should be. Our inquiry was never about fundamental ontology (a question we
can leave to a final theory, if there ever is one), but with good-enough
ontology, in a definite regime.

In the regime we are concerned with, stable particles of ordinary matter
whose number is conserved in time, there is the equivalence between
one-particle states and modes of quantum field already mentioned. Let us
settle on a preferred decomposition of the field (or preferred basis) in a
given context. But suppose that context involves non-trivial entanglement:
can entanglements of particles be understood as entanglements of modes of
fields?

Surely they can -- but on pain of introducing many more modes of the field
then there were particles, and a variable number to boot. As with
one-particle states so modes of the field:\ in a general entanglement,
arbitrarily many such modes are involved, even given a preferred
decomposition of the field, whereas the number of particles is determinate.
Just where the particle concept is the most stable, in the regime in which
particle number is conserved, eliminativism in favour of fields and modes of
fields introduces those very features of the particle concept that we found
unsatisfactory in strongly-interacting regimes. That speaks against
eliminativism.

This does not, of course, militate against the reality of quantum fields. We
recognize that permutable particles are emergent from quantum fields, just
as non-permutable particles are emergent from permutable ones. Understood in
this way, we can explain a remaining fragment of the Gibbs paradox -- the
fact that particle identity, and with it permutation symmetry, can ever be
exact. How is it that intrinsic quantities, like charge and mass, are
identically the same? (their values are real numbers, note). The answer is
that for a given particles species, the particles are one and all
excitations of a single quantum field -- whereupon these numerical
identities are forced, and permutation symmetry has to obtain. The existence
of\textit{\ exact} permutation symmetry, in regimes in which particle
equations are \textit{approximately} valid, is therefore explained, and with
it particle indistinguishability.\bigskip

\newpage

\begin{center}
{\LARGE REFERENCES\bigskip }
\end{center}

\setlength{\parindent}{-0.7cm} Albert, D. [2000], \textit{Time and Chance}. Cambridge: Harvard
University Press.

\setlength{\parindent}{-0.7cm} Bacciagaluppi, G. (2013), 'Measurement and classical regime in quantum mechanics', in  \textit{The Oxford Handbook of Philosophy of Physics}, R. Batterman (ed.), Oxford University Press. Available online at http://philsci-archive.pitt.edu/8770/

\setlength{\parindent}{-0.7cm} Bach, A. [1990], `Boltzmann's probability distribution of 1877', 
\textit{Archive for the History of the Exact Sciences} \textbf{41}\textit{,}
1-40.

--- [1997], \textit{Indistinguishable Classical Particles}.
Berlin: Springer.

\setlength{\parindent}{-0.7cm} Belot, G. (2013), `Symmetry and equivalence', in \textit{The Oxford Handbook of Philosophy of Physics}, R. Batterman (ed.), Oxford University Press.

\setlength{\parindent}{-0.7cm} Brown, H., E. Sj\"{o}qvist, and G. Bacciagaluppi (1999), `Remarks on identical particles in de Broglie Bohm theory', \textit{Physics Letters A} 251, 229-235.

\setlength{\parindent}{-0.7cm} Darrigol, O. [1991], `Statistics and cominbatorics in early
quantum theory, II: Early symptoma of indistinghability and holism', \textit{%
Historical Studies in the Physical and Biological Sciences} \textbf{21},
237-298.

\setlength{\parindent}{-0.7cm} Denbigh, K. and M. Redhead [1989], `Gibbs` paradox and non-uniform
convergence', \textit{Synthese} \textbf{81}, 283-312.

\setlength{\parindent}{-0.7cm} Dieks, D. and A. Lubberdink [2010], `How classical particles
emerge from the quantum world', \textit{Foundations of Physics} \textbf{41}, 1051-1064.
Available online at http://xxx.lanl.gov/abs/1002.2544.

\setlength{\parindent}{-0.7cm} Ehrenfest, P., and V. Trkal [1920], `Deduction of the dissociation
equilibrium from the theory of quanta and a calculation of the chemical
constant based on this', \textit{Proceedings of the Amsterdam Academy} 
\textbf{23}, 162-183. Reprinted in P. Bush, (ed.), \textit{P. Ehrenfest,
Collected Scientific Papers}, North-Holland, 1959.

\setlength{\parindent}{-0.7cm} French, S. and D. Krause [2006], \textit{Identity in physics : a
historical, philosophical, and formal analysis}, Oxford University Press.

\setlength{\parindent}{-0.7cm} French, S. and M. Redhead [1988], `Quantum physics and the
identity of indiscernibles', \textit{British Journal for the Philosophy of
Science} \textbf{39}, 233-246.

\setlength{\parindent}{-0.7cm} Fujita, S. [1990], `On the indistinguishability of classical particles', \textit{Foundations of Physics} \textbf{21}, 439-57.

\setlength{\parindent}{-0.7cm} Ghirardi, G., L. Marinatto, and Y. Weber [2002], `Entanglement and
properties of compositie quantum systems: a conceptual and mathematical
analysis', \textit{Journal of Statistical Physics} \textbf{108}, 49-122. Available online at http://arxiv.org/abs/quant-ph/0206021

\setlength{\parindent}{-0.7cm} Ghirardi, G. and L. Marinatto [2004], `General criterion for the entanglement of two indistinguishable states’, \textit{Physical Review} \textbf{A 70} 012109. Available online at http://arxiv.org/abs/quant-ph/0410086

\setlength{\parindent}{-0.7cm} Gibbs, J. W. \ [1902], \textit{Elementary Principles in
Statistical Mechanics}, Yale University Press.

\setlength{\parindent}{-0.7cm} Goldstein, S., J. Taylor, R. Tumulka, N. Zanghi [2005a], `Are all
particles identical?', \textit{Journal of Physics} \textbf{A38}, 1567-1576. Available online at http://xxx.lanl.gov/abs/quant-ph/0405039.

--- [2005b], `Are all
particles real?', \textit{Studies in the History and Philosophy of Modern
Physics }\textbf{36},103-112. Available online at http://arxiv.org/abs/quant-ph/0404134

\setlength{\parindent}{-0.7cm} Gottesman, D. [2005], `Quantum statistics with classical
particles'. Available online at http://xxx.lanl.gov/abs/cond-mat/0511207.

\setlength{\parindent}{-0.7cm} Greiner, W. and B. M\"{u}ller [1994], \textit{Quantum Mechanics:
Symmetries}, 2nd Ed., Springer.

 \setlength{\parindent}{-0.7cm} Hercus, E. [1950], \textit{Elements of Thermodynamics and
Statistical Thermodynamics}, Melbourne University Press.

\setlength{\parindent}{-0.7cm} Hilbert, D., and P. Bernays [1934], \textit{Grundlagen der
Mathematik}, Vol.1, Springer: Berlin.

\setlength{\parindent}{-0.7cm} Howard, D. [1990], `"Nicht sein kann was nicht kein darf", or the
prehistory of EPR, 1909-1935: Einstein's early worries about the quantum
mechanics of compound systems', in \textit{Sixty-Two Years of Uncertainty:
Historical, Philosophical and Physical Inquiries into the Foundations of
Quantum Mechanics}, A. Miller (ed.), Plenum.

\setlength{\parindent}{-0.7cm} Huggett, N. [1999a], `Atomic metaphysics',\textit{\ Journal of
Philosophy }\textbf{96}, 5-24.

--- [1999b], \textit{Space from Zeno to Einstein:\
Classical Readings with a Contemporary Commentary}: Bradford Books.

\setlength{\parindent}{-0.7cm} Jammer, M. [1966], \textit{The Conceptual Development of Quantum
Mechanics}, McGraw-Hill.

\setlength{\parindent}{-0.7cm} Jaynes, E. [1992], `The Gibbs paradox', in \textit{Maximum-Entropy
and Bayesian Methods}, G. Erickson, P. Neudorfer, and C. R. Smith (eds.),
Kluwer, Dordrecht.

\setlength{\parindent}{-0.7cm} van Kampen, N. [1984], `The Gibbs paradox', \textit{Essays in
theoretical physics in honour of Dirk ter Haar,} W.E. Parry, (ed.). Oxford:
Pergamon Press.

\setlength{\parindent}{-0.7cm} Leinaas, J., and J. Myrheim [1977], `On the theory of identical
particles', \textit{Il Nuovo Cimento} \textbf{37B}, 1-23.

\setlength{\parindent}{-0.7cm} Lieb, E., and J. Yngvason, [1999], `The physics and mathematics of
the second law of thermodynamics',\textit{\ Physics Reports} \textbf{310},
1-96.

\setlength{\parindent}{-0.7cm} Muller, F. and S.Saunders [2008], `Distinguishing fermions', 
\textit{British Journal for the Philosophy of Science} \textbf{59}, 499-548.

\setlength{\parindent}{-0.7cm} Muller, F. and M. Seevink [2009], `Discerning elementary
particles', \textit{Philosophy of Science} \textbf{76}, 179 - 200. Available
online at http://xxx.lanl.gov/abs/0905.3273

\setlength{\parindent}{-0.7cm} Nagle, J. [2004], `Regarding the entropy of distinguishable
particles', \textit{Journal of Statistical Physics} \textbf{117}, 1047-62.

\setlength{\parindent}{-0.7cm} Penrose, R. [2004], \textit{The Road to Reality}, Vintage Press.

\setlength{\parindent}{-0.7cm} Planck, M. [1900], `Zur Theorie des Gesetzes der Energieverteilung
im Normalspectrum', \textit{Verhandlungen der Deutsche Physicalishe Gesetzen}
\textbf{2}, 202-204. Translated in D. ter Haar (ed.), \textit{The Old
Quantum Theory}, Pergamon Press, 1967.

--- [1912], `La loi du rayonnement noir et l'hypoth\`{e}se
des quantit\'{e}s \'{e}l\'{e}mentaires d'action', in P. Langevin and M. de
Broglie (eds.), \textit{La Th\'{e}orie du Rayonnement et les Quanta -
Rapports et Discussions de la R\'{e}sunion Tenue \`{a} Bruxelles, 1911},
Gauthier-Villars.

--- [1921], \textit{Theorie der W\"{a}rmesrahlung. }4th
ed., Leipzig: Barth, Leipzig.

\setlength{\parindent}{-0.7cm} Pniower, J. [2006], \textit{Particles, Objects, and Physics}, D.
Phil Thesis, University of Oxford. Available online at http://philsci-archive.pitt.edu/3135/

\setlength{\parindent}{-0.7cm} Poincar\'{e}, H. [1911], `Sur la theorie des quanta', \textit{%
Comptes Rendues} \textbf{153}, 1103-1108.

--- [1912], `Sur la theorie des quanta', \textit{%
Journal de Physique} \textbf{2}\textit{,} 1-34.

\setlength{\parindent}{-0.7cm} Pooley, O. [2006], `Points, particles, and structural realism', \
in D. Rickles, S. French, and J. Saatsi (eds.),\textit{The Structural
Foundations of Quantum Gravity}, Oxford University Press.

\setlength{\parindent}{-0.7cm} Post, H. [1963], `Individuality and physics', \textit{The Listener}%
, 10 October, pp. 534-537, reprinted in \textit{Vedanta for East and West} 
\textbf{132}, 14-22 (1973).

\setlength{\parindent}{-0.7cm} Quine, W. van [1960], \textit{Word and Object}, Cambridge: Harvard
University Press.

--- [1970], \textit{Philosophy of Logic}, Cambridge:
Harvard University Press.

--- [1990], \textit{The Pursuit of Truth}, Cambridge:
Harvard University Press.

\setlength{\parindent}{-0.7cm} Rapp, D. [1972], \textit{Statistical Mechanics}. New York: Holt,
Rinehart and Winston.

\setlength{\parindent}{-0.7cm} Reichenbach, H. [1956], \textit{The Direction of Time}, University
of California Press.

\setlength{\parindent}{-0.7cm} Rosenfeld, L. [1959], `Max Planck et la definition statistique de
l'entropie', \textit{Max-Planck Festschrift 1958}, Berlin: Deutsche Verlag
der Wissenschaften. Trans. as `Max Planck and the statistical definition of
entropy', in R. Cohen and J. Stachel (eds.), \textit{Selected Papers of Leon
Rosenfeld}, Reidel, 1979.

\setlength{\parindent}{-0.7cm} Saunders, S. [1991], `The negative energy sea', in \textit{%
Philosophy of Vacuum}, S. Saunders and H. Brown (eds.), Oxford: Clarendon
Press.

--- [1992], `Locality, complex numbers, and relativistic
quantum theory', \textit{Proceedings of the Philosophy of Science
Association Vol.1}.

--- [2003], `Physics and Leibniz's principles', in K.
Brading and E. Castellani (eds.]), \textit{Symmetries in Physics: New
Reflections}. Cambridge University Press. Available online at http://philsci-archive.pitt.edu/2012/

--- [2006a], `On the explanation of quantum statistics', 
\textit{Studies in the History and Philosophy of Modern Physics} \textbf{37}%
, 192-211. Available online at http://arxiv.org/abs/quant-ph/0511136

--- [2006b], `Are quantum particles objects?', \textit{%
Analysis} \textbf{66}, 52-63. Available online at
philsci-archive.pitt.edu/2623/

--- [2007], `Mirroring as an a priori symmetry', \textit{%
Philosophy of Science} \textbf{74}, 452-480. Available online at http://philsci-archive.pitt.edu/3809/

--- (2016), `The emergence of individuals in physics', in \textit{Individuals Accross the Sciences}, A. Guay and T. Pradeu (eds.), Oxford University Press. 

\setlength{\parindent}{-0.7cm} Schr\"{o}dinger E. [1946], \textit{Statistical Thermodynamics}.
Cambridge: Cambridge University Press.

--- [1984], `What is an elementary particle?', 
\textit{Collected Papers, Vol.4}, \"{O}sterreichische Akademie der
Wissenschaften. Reprinted in \textit{Interpreting Bodies: classical and
quantum objects in modern physics}, E. Castellani (ed.), Princeton
University Press, 1998.

\setlength{\parindent}{-0.7cm}  Shankar, R. [1980], \textit{Principles of Quantum Mechanics}, Yale
University Press.

\setlength{\parindent}{-0.7cm} Stern, O. [1949], `On the term $N!$ in the entropy', \textit{Reviews
of Modern Physics} \textbf{21}, 534-35.

\setlength{\parindent}{-0.7cm} Swendsen, R. [2002], `Statistical mechanics of classical systems
with distinguishable particles', \textit{Journal of Statistical Physics} 
\textbf{107}, 1143-66.

--- [2006], `Statistical mechanics of colloids and
Boltzmann's definition of the entropy', \textit{American Journal of Physics} 
\textbf{74}, 187-190.

\setlength{\parindent}{-0.7cm} Teller, P. [1995], \textit{An Interpretative Introduction to
Quantum Field Theory}, Princeton University Press.

\setlength{\parindent}{-0.7cm} Wallace, D. [2013], `The Everett interpretation', in \textit{The Oxford Handbook of Philosophy of Physics}, R. Batterman (ed.), Oxford University Press. Available online at http://philsci-archive.pitt.edu/8888/

\setlength{\parindent}{-0.7cm}Wiggins, D. [2004], \textit{Sameness and Substance Renewed},
Oxford: Oxford University Press.

\end{document}